\documentclass[aps,prb,reprint,superscriptaddress,twocolumn,10pt]{revtex4-2}

\usepackage{ragged2e}
\usepackage{graphicx}   
\usepackage{wrapfig}    
\usepackage{float}      
\usepackage{bm}
\usepackage[usenames,dvipsnames,svgnames,table]{xcolor}
\usepackage{epsfig}
\usepackage{amsmath}

\usepackage{tikz}
\usetikzlibrary{calc} 
\usepackage{pgfplots}

\usepackage{amssymb}
\usepackage{textgreek}
\usepackage{array}
\usepackage{nccmath}
\usepackage{dsfont}
\usepackage{array}
\usepackage{cancel}
\usepackage{comment}
\usepackage{hyperref}
\usepackage{cleveref}
\crefname{figure}{Fig.}{Figs.}
\hypersetup{colorlinks=true,breaklinks,linkcolor=blue,urlcolor=blue,citecolor=blue}
\usepackage{orcidlink}
\usepackage{algpseudocode}
\usepackage{soul}
\usepackage[normalem]{ulem}

\usepackage{booktabs}

\definecolor{fitblue}{HTML}{1F77B4}
\definecolor{fitorange}{HTML}{FF7F0E}
\definecolor{fitgreen}{HTML}{2CA02C}
\definecolor{fitred}{HTML}{D62728}
\definecolor{fitkhaki}{HTML}{BCBD22}
\definecolor{fitpurple}{HTML}{9467BD}

\allowdisplaybreaks

\newcommand{\fbc}[1]{{\color{black} #1}}

\newcommand{\fbt}[1]{{\color{black} #1}} 

\newcommand{\gbc}[1]{{\color{black} #1}}

\newcommand{\kh}[1]{{\color{black} #1}}

\newcommand{\khx}[1]{{\color{olive} }}

\definecolor{darkred}{rgb}{0.55,0,0}

\DeclareRobustCommand{\khca}[1]{{\color{black} #1}} 
\DeclareRobustCommand{\khcax}[1]{{\color{darkred} }}
\newcommand{\khcac}[2][]{\if\relax\detokenize{#1}\relax\else\label{#1}\fi{\color{darkred}[{\bf KH:} #2]}} 

\begin{document}
\title{Scaling of entanglement with distance between subsystems: \\ detection of phase transitions in the $t'$-$\pi$-flux model}

\author{Maksim Ulybyshev \orcidlink{0000-0003-3135-5704}} 
\affiliation{%
    Institut für Theoretische Physik und Astrophysik, Universität W\"urzburg, 97074 W\"urzburg Germany }%

\author{Gerg\H o Ro\'osz \orcidlink{0000-0002-9541-119X}} 
\affiliation{%
    Institute of Solid State Physics, TU Wien, 1040 Vienna, Austria}%

\author{Gabriele Bellomia \orcidlink{0000-0003-1576-3388}}
\affiliation{%
    Institute of Solid State Physics, TU Wien, 1040 Vienna, Austria}%

\author{Karsten Held \orcidlink{0000-0001-5984-8549}}
\affiliation{%
    Institute of Solid State Physics, TU Wien, 1040  Vienna, Austria}%

\author{Fakher Assaad \orcidlink{0000-0002-3302-9243}} 
\affiliation{%
    Institut für Theoretische Physik und Astrophysik, Universität W\"urzburg, 97074 W\"urzburg Germany }%

 \affiliation{%
    W\"urzburg-Dresden Cluster of Excellence ctd.qmat, Am Hubland, 97074 W\"urzburg, Germany}%

\author{Frederic Bippus \orcidlink{0009-0006-4316-6547}}
\affiliation{%
    Institute of Solid State Physics, TU Wien, 1040 Vienna, Austria}%
\affiliation{%
    Center for Computational Quantum Physics, Flatiron Institute, New York, NY 10010}

\date{\today}
    
\begin{abstract}
The system\kh{-}size scaling of bipartite entanglement 
can be used to detect phase transitions. 
We \kh{develop a related approach} \kh{for} strongly correlated electron models where large density 
matrices {or large system size scaling} are inaccessible. To this end,
we demonstrate that phase transitions of various kinds are signaled by a 
sharp signature in the first- or second-order logarithmic derivatives of the negativity measuring entanglement 
between small clusters at  varying distances. Crucially, we show how to mitigate the effect of quantum Monte Carlo noise at large distances when computing  the negativity.
Our focus is on \kh{the \kh{$t'$-$\pi$-f}lux Hubbard model using auxiliary\khca{-}field quantum Monte Carlo simulations.}
\kh{It has the particular advantage of showing  three distinct phase transitions in a single model:} 
semimetal to band insulator, semimetal to \kh{antiferromagnetic} Mott insulator,  and a magnetic order-disorder transition
belonging to the Gaussian, Gross-Neveu, and O(3) Wilson-Fisher universality classes, respectively.
\end{abstract}
\maketitle


\section{Introduction}
A key challenge in condensed matter theory is the detection and 
classification of phase transitions 
when the Ginzburg–Landau
theory for first- and second-order phase
transitions \cite{Kopietz2010} is not applicable. Order-parameter-free 
approaches can be based, among others, on the fidelity (ground-state overlap) \cite{Zanardi_2006,Gu_2010_fidelity}, machine learning \cite{Carrasquilla_2017,vanNieuwenburg_2017,Broecker17} or tensor network properties 
\cite{rohshap_QTTD_2026},
and, most prominently, on entanglement. For the latter, the quantum Fisher 
information \cite{hauke_measuring_2016} can be 
used  \cite{Yamashika_QFI_symbreaking_2025}\kh{, and has been used for detecting} quantum criticality 
\cite{fang_amplified_2025,Mathew2020,mazza_quantum_2024,laurell_quantifying_2021,Scheie2024,wuttig2026} \kh{as well as} multipartite entanglement in cuprates
\cite{Bippus_2025_pseudogap_QFI,balut2025quantumfisherinformationreveals,tenhuisen2026}.
\kh{Even more} \kh{often, at least for theoretical calculations,} bipartite entanglement measures are considered \cite{amico_entanglement_2008}. For example, a 
phase transition can be detected through the system\khca{-}size scaling of the entanglement 
entropy \cite{amico_entanglement_2008,Eisert_2010,islam_measuring_2015,Tagliacozzo_2008,Pollmann_2009}, an approach 
that can be extended to finite-temperature entanglement measures
\cite{Wilms_2012,Iaconis_2013}.

These approaches, however, fail in strongly interacting fermionic models where exact
solutions are rare and computations are limited to small system sizes. 
In addition, even when a numerical calculation of physically relevant operators 
for relatively large lattices is feasible, obtaining and storing the
reduced density matrix of an extensive subsystem poses yet another hard numerical problem.
Luckily, for some transitions entanglement between neighboring small subsystems already shows clear signatures: in spin systems \cite{osterloh_scaling_2002,Osborne_2002,Lu_2019}, the Kitaev honeycomb model \cite{Cui_2010}, 
in one- and two-impurity Kondo systems \cite{Dong_2021,LoSchiavo2026}, and {at} the opening of
a Mott gap \cite{Bellomia2023} 
 or pseudogap \cite{BellomiaPhD,bippus2026_qsim} 
in the \emph{plain\kh{-vanilla}} 
two-dimensional \kh{(2D)} Hubbard model.
However, this \kh{need} not necessarily be the case.

\begin{figure}
    \includegraphics[width=1\linewidth]{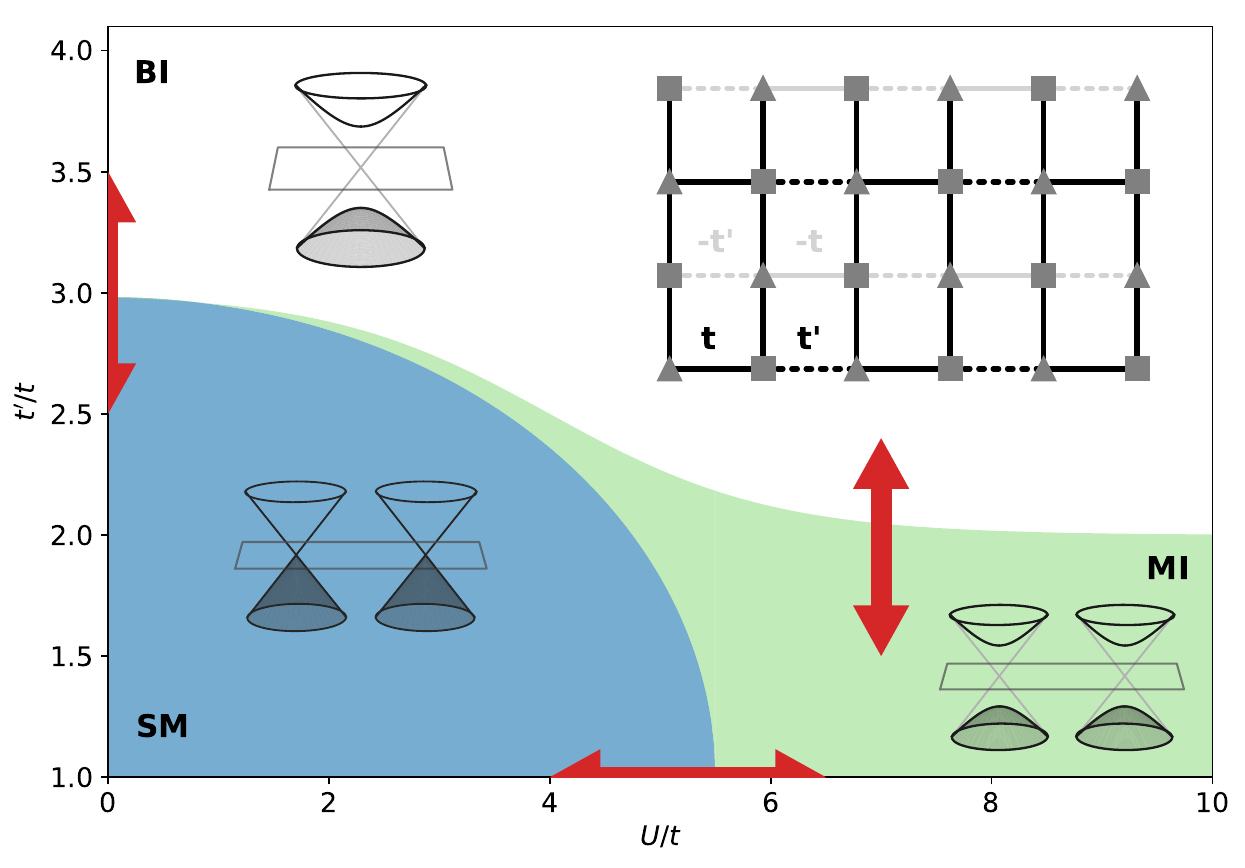}
    \caption{Schematic phase diagram of the $t'$-$\pi$-flux Hubbard model \cite{Ixert_piflux_2014}.
    The inset shows the Dirac-point bandstructure of the three phases\khcax{;}\khca{} and
    \kh{the hopping and} sublattice structure \fbc{ (triangles and squares)}. In the \emph{SM} regime, the band-structure forms 
    two Dirac cones\khcax{, at half-filling}\khca{;  at half filling} (\khca{gray} shading) the Fermi surface 
    is made up from the singular points of the two cones.
    Once a $t'$ is introduced, the cones move through momentum space until 
    merging in the \emph{BI} regime with a non-interacting bandgap opening, whereas 
    for the \kh{antiferromagnetic} \emph{MI} regime, the non-interacting band-structure remains unchanged but interactions open a Mott \kh{or Slater} gap. Red arrows indicate the transitions we investigate in this paper. }
    \label{fig:phase_digram}
\end{figure}

Without varying the simulated system size and remaining at small partition sizes, we therefore investigate if 
a phase transition can be  detected by increasing the distance between these
small subsystems. We call this
approach ``scaling with distance'' in the following.

Small subsystems at a finite distance have been considered before, but with different scaling variables.
Correlations between two distant clusters, captured by the correlation density matrix \cite{Cheong_2009},
were recently sampled with QMC and  analyzed by finite-size scaling at maximal separation \cite{Chincholi_2025}.
The negativity between separated regions has been studied as a function of distance at criticality \cite{Calabrese_2012,Parez_2024_separated}, including with QMC in the $\pi$-flux
$t$-$V$ model \cite{wang_entanglement_2025}. 
In one-dimensional free fermions, the negativity between
separated blocks distinguishes symmetry-protected topological phases and topologically distinct critical
points \cite{Choi2024}.
Entanglement between small subsystems has also been used to construct measures of
genuine multipartite entanglement, employed to detect phase transitions in the quantum
Ising model \cite{Lyu_2025}, and recently extended to fermions and studied as a function
of temperature and separation in the Kitaev chain \cite{Allen_2026}.
In all these works, the separation is either kept fixed or tied to the system size, or varied at a fixed coupling, typically at the critical point to extract universal exponents.
Here, we instead follow the coupling dependence of the negativity at a sequence of distances $d$, so that
$d$ plays a role analogous to that of the system size.
This is motivated, for example, by the distance scaling of the
negativity observed in Ising chains with defects \cite{rogerson_entanglement_2022}.

To test if phase transitions can indeed be resolved by entanglement at large distances we investigate the rich phase diagram of the $t'$-$\pi$-flux Hubbard model (Fig.~\ref{fig:phase_digram}). It starts from the usual $\pi$-flux model on the square lattice, featuring two Dirac cones in the semimetal (\emph{SM}) phase at low interaction strength $U$. Once the repulsive Hubbard interaction increases, at a critical coupling of around $U_\mathrm{c}\approx 5.5$, the model undergoes a Gross-Neveu phase transition  \cite{Herbut09a,Assaad13,ParisenToldin_2015,Otsuka_2016,Chang_2012,Wang26} to an antiferromagnetic (AFM) Mott insulator (\emph{MI}). This transition is shown along the $t'/t=1$ line in Fig.~\ref{fig:phase_digram}. Subsequently, we add a modulation of hoppings in the horizontal direction, introducing $t'\neq t$, as shown in the inset of Fig.~\ref{fig:phase_digram}. Once $t'$ increases, the two Dirac cones move towards each other in the Brillouin zone (BZ), eventually colliding (at $t'=3$ in the non-interacting case) and disappearing \cite{Ixert_piflux_2014,Montambaux_2009,Dietl_2008,Tarruell_2012}. This way, the system becomes a band insulator (\emph{BI}), even at zero Hubbard interaction. Finally, at large interaction $U>U_\mathrm{c}$, we have a direct transition from the \emph{MI} to the \emph{BI} phase by increasing $t'$ while keeping $U$ constant. This magnetic order-disorder  transition belongs to  the 2+1D O(3) universality class \cite{Vicari02}.

We begin by investigating the non-interacting \emph{SM-BI} transition
where large distances can be reached and results can be compared 
with the ground-state entanglement entropy of extended systems \cite{baggioli2023}. 
As a bipartite entanglement measure, we apply the fermionic negativity 
\cite{Shapourian_2017_first_NF,Shapourian_2019_second_NF} which extends the well\khca{-}established positive partial transpose criterion (PPT) \cite{Horodecki_1996,Peres_1996} to fermionic modes. Here, the existence of local-parity violating matrix
elements ensures the existence of fermionic entanglement \cite{Benatti_2014}. 
For this application, this is beneficial over parity and charge superselected entanglement (PSSR and NSSR) 
\cite{bippus2026_qsim,Banhuls2007,Friis2013,Friis2016,Vidal2021,Benatti2020,Szalay_2021,DelMaestro_accessible2019,DelMaestro_accessible2024,Ding_2024,Ding2022QuantifyingEE, Bellomia2023, BellomiaPhD,Wick_charge_SSR,Wick_parity_SSR}, which we will discuss below in more detail, since it suffers no sudden death, neither with temperature \cite{thermal_death_SSR,Parez_2024_fate} nor with distance \cite{Parez_2024_fate}.

We then extend our approach to the interacting \emph{MI-BI} transition.
To access this regime, we apply a finite-temperature auxiliary-field quantum Monte Carlo 
solver (QMC) \cite{Blankenbecler81,White89,Assaad08_rev} along the lines of the ALF-code \cite{ALF_v2.4}. On the technical side,
building on earlier reconstructions of few-site reduced density matrices
from Green's functions \cite{PhysRevB.110.075115} and in determinant QMC \cite{Humeniuk_2019,wang_entanglement_2025},
we introduce a general methodology to compute 
arbitrary equal-time and time-displaced correlators from QMC which enables the efficient computation of multi-site reduced density matrices.  This allows for the computation of entanglement measures, beyond the often considered Rényi entropy \cite{Grover_2013,Assaad_2014,Broecker_2014} and Rényi negativity \cite{Wang25}.  To this end, we also show how to extract the negativity from noisy data, including QMC results, by extending Vidal et al.'s formula for the negativity \cite{Vidal_2002_logN}.  The idea behind the method to compute these density matrices from QMC is a computer algebra of linear operators acting on vectors in the Fock space, realized directly in the QMC code \cite{superlat_github}. Thus the observables can be simply written as expressions in terms of fermionic creation and annihilation operators. Subsequently, these expressions are automatically translated into products of fermionic Green's functions, averaged over configurations of auxiliary fields in the QMC process. In addition to that, we also select, for each of the observables needed for the reduced density matrix, the Hubbard-Stratonovich channel (spin or charge) with the smaller fluctuations.  Details of the algorithms are discussed in Sec.~\ref{sec:QMC}.

Finally, we consider the \emph{SM-MI} transition. The opening of a Mott gap is
interaction driven and belongs to the class of Gross-Neveu transitions for 
which quantum criticality was detected by the entanglement entropy in the 
$\pi$-flux model \cite{jiang_uncovering_2026,liao_universal_2025} and by
the fermionic negativity in the spinless $t$-$V$ model on the $\pi$-flux lattice \cite{wang_entanglement_2025}.
In the latter case, the nearest-neighbor negativity was found to be only weakly sensitive to the transition,
which further motivates going to larger separations.


\section{$t'$-$\pi$-flux model}
The \kh{$t'$-$\pi$-flux}  model features \kh{three phases: \kh{s}emimetal, band\khcax{-} insulator and antiferromagnetic (AFM) Mott insulator with phase transitions between them.}
\kh{Its} Hamiltonian on the square lattice reads:
\begin{equation}
    \hat H = - \sum_{<\vec x, \vec y>, \sigma}  t_{\vec x,\vec y} \hat c^\dagger_{\vec x,\sigma} \hat c_{\vec y, \sigma} + \mathrm{h.c.} + U \sum_{\vec x} \hat n_{\vec x, \uparrow} \hat n_{\vec x, \downarrow}, \label{eq:basic_ham}
\end{equation}
  where $\hat c^\dagger_{\vec x,\sigma}$ is the creation operator for
  an electron with spin $\sigma=\uparrow, \downarrow$ on site $\vec x=(x_1, x_2)$\kh{;}
  $n_{\vec x, \sigma}= \hat c^\dag_{\vec x,\sigma} \hat c_{\vec x, \sigma}$; \kh{$\langle ... \rangle$ restricts the first sum to nearest neighbors (each pair summed once). } 
   
Hoppings are distributed according to the inset in  Fig.~\ref{fig:phase_digram}, and $t=1$ sets all energy scales. At $t=t'$,
the model reduces back to the usual $\pi$-flux model\kh{, with alternating signs
of $t$ along the $y$-direction}. However, if $t'$ is larger than $t$, Dirac points
start to move towards each other and finally collide at $t'=3t$. If $t'>3t$, 
the model is a band  insulator. 
We confirm the parameters for th\kh{is and the other} transitions in Appendix~\ref{Appendix_phase_transitions}.


\section{Entanglement Measures}
We consider a density matrix $\rho_{AB}$ on two subsystems $A$ \kh{and} $B$, 
defined by tracing out the complement of the \kh{union of the}  subsystems \kh{$A$ and $B$:}
\begin{equation}\label{eq_red_rho}
    \rho_{AB} = \textrm{tr}_{\overline{AB}} \rho.
\end{equation}
Specifically, we consider either \kh{$A$ and $B$ both consisting of a}
pair of sites to measure \emph{dimer-dimer} entanglement,  or 
\kh{consisting of one site each only} to measure the \emph{two-site / inter-site} (2s) entanglement.

Such a  state is separable 
if it can be written as a classical combination of product states
\begin{equation}\label{seperabel_rho}
\rho_{AB} = \sum_{n} p_n \; \rho^{n}_A \otimes \rho^{n}_B\kh{,}
\end{equation}
\kh{with a probability distribution $p_n$.}
Otherwise, it is entangled. Such entanglement is detected as described \kh{in the following}.

\subsection{Fermionic Negativity}

The fermionic negativity \cite{Shapourian_2017_first_NF,Shapourian_2019_second_NF} ${\cal N}^\mathrm{F}$ is an extension of the well\khca{-}established bosonic negativity 
\cite{Peres_1996,Horodecki_1996,Bippus_2sRDM_entanglement_2026},
which is based on the positive semi-definiteness of any physical 
density matrix. This property is preserved under partial transpose of
\kh{one subsystem (say $B$)} if the density matrix is separable.
Vice versa, any violation of this property signals entanglement. 
The basic partial transpose\khca{}
acts on an operator $O$ as 
\begin{equation}\label{eq:PT}
    \langle \psi_A, \psi_B | O^{T_B} | \bar{\psi}_A, \bar{\psi}_B \rangle \equiv  \langle \psi_A,  \bar{\psi}_B |O | \bar{\psi}_A, \psi_B \rangle.
\end{equation}
To reinstate the fermionic signs that are missed by this operation,
the \kh{\it fermionic} negativity replaces the partial transpose by
the partial time-reversal operation
\begin{equation}\label{eq:fermionic_transpose}
    \langle \psi_A, \psi_B | O^{{\tau}_B} | \bar{\psi}_A, \bar{\psi}_B \rangle \equiv  \langle \psi_A,  \bar{\psi}_B |O | \bar{\psi}_A, \psi_B \rangle (-1)^{\phi}.
\end{equation}
Here, the phase factor \kh{$\phi$} is given in terms of the 
number of electrons $n_{A(B)} = \sum_{i\in A(B)}\khca{(}n_{i,\uparrow} + n_{i,\downarrow}\khca{)}$ 
of each component $\psi_{A(B)},\bar\psi_{A(B)}$ of the density matrix as 
\begin{equation}
\begin{aligned}
\phi = & \frac{n_A\left(n_A+2\right)}{2}+\frac{\bar{n}_A\left(\bar{n}_A+2\right)}{2}+\bar{n}_B n_B + n_A n_B+\bar{n}_A \bar{n}_B \\
&+\left(\bar{n}_A+\bar{n}_B\right)\left(n_A+n_B\right).
\end{aligned}
\end{equation}

Entanglement is measured via this operation by
\begin{equation}\label{eq:fermionic_negativity}
    {\cal N}^{\mathrm{F}}(\rho_{AB}) \equiv \frac{\| \rho_{AB}^{\tau_B} \|_{\textrm{tr}}-1}{2} = \frac{\| \rho_{AB}^{\tau_B} \|_{\textrm{tr}}-\mathrm{Tr}\rho_{AB} }{2}
\end{equation}
with $\| \rho \|_{\textrm{tr}} \equiv \textrm{tr}\sqrt{(\rho \rho^{\dagger})}$ \cite{Shapourian_2017_first_NF,Shapourian_2019_second_NF,Vidal_2002_logN}.
Note that due to the noisy character of QMC data, we need to use the rightmost expression in order to correct for small fluctuations of the trace at large $d$.
Specifically, we will plot the logarithmic negativity  \cite{Vidal_2002_logN,Plenio_2005_logN}
\begin{equation}
    N^\mathrm{F} \equiv \log_2\left( 2{\cal N}^{\mathrm F}+1 \right).
\end{equation}
Any value $N^\mathrm{F}>0$ detects entanglement while $N^\mathrm{F}=0$ is inconclusive. 
This entanglement however can be due to locally inaccessible states \cite{Wick_charge_SSR,Wick_parity_SSR}. From a quantum information perspective, it may therefore be favorable to restrict the density matrix to the superselected sector \cite{Friis2013,Friis2016,Banhuls2007}.

\subsection{Spin entanglement} 
In the Gross-Neveu transition from semi-metal to Mott insulator,
charges are gradually localized and at large values of interaction 
$U$ only spin\khcax{-} fluctuations remain. This motivates us to add a new
angle of understanding to this process by differentiating between 
entanglement purely from spin fluctuations and \kh{the entanglement measure of the previous subsection} that includes charge fluctuations. 
To this end, we propose the following filtering
\begin{equation}
    \rho_{AB}^{\textrm{S}} =\!\!\!\sum_{{\kh{N_{i\in A} N_{j\in B}}}}  \prod_{i\in A} P_{N_i} \kh{\otimes} \prod_{j\in \kh{B}} P_{N_j} \;  \rho_{AB} \;  \prod_{i\in A} P_{N_i}  \kh{\otimes}\prod_{j\in \kh{B}} P_{N_j},
    \label{eq:sel_rules}
\end{equation}
where we eliminate all
off-diagonal components of the density matrix that do not 
preserve the local charge \kh{$N_{i(j)}$  on each and every
lattice site $i(j)\in A(B)$}. 
Note that this deletes any quantum coherence that introduces
phase factors into the partially time-reversed density matrix Eq.~\eqref{eq:fermionic_transpose},
such that it collapses back onto the regular (bosonic) partial transpose Eq.~\eqref{eq:PT}. Here, $P_{N_\kh{l}}$ are the
projectors onto subspaces with constant charge
\begin{equation}
    N_\kh{l} = n_{l\uparrow} + n_{l\downarrow}
\end{equation}
on lattice site $l$.
When applied to one-site subsystems of a two-site reduced
density matrix, this is of course nothing more than the 
well-established fermionic charge superselection rule\khcax{.}\khca{}

\cite{Banhuls2007,Friis2013,Friis2016,Vidal2021,Benatti2020,Szalay_2021,DelMaestro_accessible2019,DelMaestro_accessible2024,Ding_2024,Ding2022QuantifyingEE, Bellomia2023, BellomiaPhD, bippus2026_qsim}. 
\kh{Let us} note that it is well justified to measure entanglement with the 
negativity from NSSR constrained density matrices \cite{thermal_death_SSR}.

However, when extending  \kh{$A$ or $B$} to larger 
subsystems\kh{,} the $\rho^\mathrm{S}_{AB}$ density matrix is
more restricted than the NSSR density matrix, {which we also introduce here for completeness}:
\begin{equation}
    \rho_{AB}^{\kh{\mathrm{NSSR}}} = \sum_{N_{A},N_{B}} P_{N_{A}} \otimes P_{N_{B}} \; \rho_{AB} \; P_{N_{A}} \otimes P_{N_{B}}\kh{,}
\end{equation}
\kh{with $N_{A(B)}=\sum_{l\in A(B)}N_l$.}
Importantly, the spin filtering deletes additional off-diagonal elements\khca{,} thereby ensuring that 
\begin{equation}
    N^{\textrm{S}} \leq N^{\textrm{NSSR}}.
\end{equation}
We formalize this connection in Appendix~\ref{SSR_vs_spin_resolved}.

It is also noteworthy that NSSR rules ensure that all
measured entanglement is in general \khcax{operational}\khca{operationally} accessible
\cite{DelMaestro_accessible2019,DelMaestro_accessible2024,Ding_2024} \kh{when measuring  only observables on 
either subsystem $A$ or $B$}. \kh{However, actual} measurement 
thereof in real solid\khca{-}state systems remains a major 
challenge\kh{, while} recent advancements in \khca{ultracold-atom} experiments 
have opened \kh{a} route to experimentally study 
entanglement \cite{bippus2026_qsim}.
\khca{However,} even these 
experimental realizations are restricted to the measurements of only
spin-resolved density correlators \cite{chalopin2025connectedcorrelationscoldatom}. 
Therefore, obtaining an NSSR filtered density matrix beyond the two-site
case remains out of reach \kh{in experiment}, while our proposed \khca{spin-filtered} density matrices are experimentally obtainable in \khca{ultracold-atom} experiments{~\cite{bippus2026_qsim}}.

\subsection{Entanglement Entropy}
To characterize the entanglement between the two halves of the system
at zero temperature at the semi\khcax{-}metal\khca{--}band\khca{-}insulator transition\kh{,} we 
\kh{additionally} use the entanglement entropy \cite{EE_black_holes, EE_local_op_1996}.
\kh{In this respect, w}e use two \khcax{kind}\khca{kinds} of cuts\kh{:} either along the  $x$  or $y$ direction where one subsystem is characterized \khcax{with}\khca{by} $y \le L/2$ or $x \le L/2$ respectively.

For the entanglement entropy we use extended subsystems of the size of  half
of the lattice\kh{, which we can calculate for $U=0$}. To avoid confusion
with the one and two site subsystems used for the other measures, we do
not label these halves by letters, instead we will denote them with the
inequality characterizing the included sites - which is $x \le L/2$ or $y \le L/2$ 
depending on the direction of the boundary between the subsystems. 

\khca{Let us} consider the reduced density matrices
\begin{align}
\rho(y \le L/2) & = \textrm{tr}_{y>L/2} \rho,  \\
\rho(x \le L/2) & = \textrm{tr}_{x>L/2} \rho.
\end{align}
\kh{Then,} the von Neumann entropy of these reduced density matrices 
defines the entanglement entropy
\begin{align}
S(y \le L/2) &= - \textrm{tr}_{y \le L/2} \rho(y \le L/2) \log \rho(y \le L/2),  \\
S(x \le L/2) &= - \textrm{tr}_{x \le L/2} \rho(x \le L/2) \log \rho(x \le L/2).
\end{align}
In this bipartite setting, the entanglement entropy provides a faithful
measure of the entanglement of the two subsystems at $T=0$.

\section{Computation of density matrices}
\label{sec:QMC}
We compute \khca{the }elements of \khca{the }density matrix following the approach developed in 
\khcax{Ref.~ }\khca{Refs.~}\cite{Fano1957, PhysRevB.110.075115}, where the density matrix was constructed
from average values of self-adjoint operators  $\hat A_i$, $i=1...d^2$, 
with $d$ being the \khcax{dimensionality}\khca{dimension} of the Hilbert space:
\begin{eqnarray}
\hat \rho = \frac{1}{ \operatorname{Tr}\!\left(e^{-\hat H\beta}\right)} \sum_i \hat A_i \operatorname{Tr}\!\left(\hat A_{i}\,e^{-\hat H\beta}\right).
\end{eqnarray}

Here, the operators $\hat A_i$ must satisfy the relations:
\begin{align}
    \hat A_i &= \hat A_i^{\dagger}, \\
    \operatorname{Tr}\, \hat A_i \hat A_j &= \delta_{ij}.
\end{align}

\khca{A common choice of basis are the} projectors onto states defined through the
occupation numbers. Let us say we want to compute the reduced 
density matrix (see Eq.~\eqref{eq_red_rho}) for $N$ sites\khcax{ ,}\khca{;} then the \khcax{occupation numbers}\khca{occupation-number} basis 
consists of $d=4^N$ vectors
\begin{eqnarray}
    |\nu_n\rangle
    =
    \bigl|
    (n_{\uparrow}^{1}\,n_{\downarrow}^{1})\,\ldots\,
    (n_{\uparrow}^{N}\,n_{\downarrow}^{N})
    \bigr\rangle,
\end{eqnarray}
where each pair $(n_{\uparrow}^{\kh{l}}\,n_{\downarrow}^{\kh{l}})$ corresponds to occupation numbers for spin up and spin down electrons at site $\kh{l}$.

Subsequently, $\hat A$ operators are written as
\begin{eqnarray}
\hat A_{i=(n,m)}=
\begin{cases}
\displaystyle
\frac{1}{\sqrt{2}}
\left(
|\nu_n\rangle\langle\nu_m|
+
|\nu_m\rangle\langle\nu_n|
\right), & n\neq m,\\[1.1em]
|\nu_n\rangle\langle\nu_n|, & n=m.
\end{cases}
\end{eqnarray}

Practically, we need to express the $\hat A_i$ operators in terms of 
fermionic creation and annihilation operators in order to 
compute $\langle \hat A_i\rangle$ in QMC. We start from the expression for $|\nu_n\rangle$:
\begin{align}
&\bigl|
(n_{\uparrow}^{1}\,n_{\downarrow}^{1})\,\ldots\,
(n_{\uparrow}^{N}\,n_{\downarrow}^{N})
\bigr\rangle \\
&\qquad=
\left(c_{1\uparrow}^{\dagger}\right)^{n_{\uparrow}^{1}}
\left(c_{1\downarrow}^{\dagger}\right)^{n_{\downarrow}^{1}}
\cdots
\left(c_{N\uparrow}^{\dagger}\right)^{n_{\uparrow}^{N}}
\left(c_{N\downarrow}^{\dagger}\right)^{n_{\downarrow}^{N}}
|0\rangle, \nonumber
\end{align}
where $|0\rangle$ is the state with all $N$ sites 
being empty. 

Thus any average $\langle \hat A_{(n,m)}\rangle$ can 
be computed using expressions like
\begin{align}
&\operatorname{Tr}\!\left(
|\nu_n\rangle\langle\nu_m|\,e^{-\beta H}
\right) \\
&=\operatorname{Tr}\!\Biggl(
\left(c_{1\uparrow}^{\dagger}\right)^{n_{\uparrow}^{1}}
\left(c_{1\downarrow}^{\dagger}\right)^{n_{\downarrow}^{1}}
\cdots
\left(c_{N\uparrow}^{\dagger}\right)^{n_{\uparrow}^{N}}
\left(c_{N\downarrow}^{\dagger}\right)^{n_{\downarrow}^{N}}
\,|0\rangle\langle 0 |   \nonumber \\
&\qquad\qquad\times
\left(c_{N\downarrow}\right)^{m_{\downarrow}^{N}}
\left(c_{N\uparrow}\right)^{m_{\uparrow}^{N}}
\cdots
\left(c_{1\downarrow}\right)^{m_{\downarrow}^{1}}
\left(c_{1\uparrow}\right)^{m_{\uparrow}^{1}}
 e^{-\beta H}  \nonumber
\Biggr).
\end{align}

\khcax{Projector to empty sites states}\khca{The projector onto the empty state} $|0 \rangle\langle 0 |$ can be written as:
\begin{align}
    P_0=| 0 \rangle\langle 0 |
    =\prod_{k=1}^{N}
    \left(1-\hat n_{\uparrow}^{k}\right)
    \left(1-\hat n_{\downarrow}^{k}\right).
\end{align}
Further simplification of expressions is possible if we 
start not from empty, but from completely filled $N$ sites:
\begin{align}
    |\Sigma\rangle
    =
    |(1,1)\,\ldots\,(1,1)\rangle.
\end{align}

\begin{align}
    \hat P_{\Sigma}
    =
    |\Sigma\rangle\langle\Sigma|
    =
    \prod_{k=1,\sigma}^{N}\hat n_{k,\sigma},  \label{eq:proj_fin}
\end{align}
where $\hat n_{k,\sigma} = c_{k\sigma}^{\dagger}c_{k\sigma},$ 
$\sigma=\uparrow,\downarrow$.  In this case the projection 
operator $\hat P_{\Sigma}$ consists of only one product, and basis vectors 
can still be written as
\begin{align}
    |\nu_n\rangle
    =
    \prod_k
    \left(\hat c_{k\uparrow}\right)^{1-n_{\uparrow}^{k}}
    \left(\hat c_{k\downarrow}\right)^{1-n_{\downarrow}^{k}}
    |\Sigma\rangle. \label{eq:vec_fin}
\end{align}

Collecting everything together, we get expressions like
\begin{align}
    \operatorname{Tr}\!\left( |\nu_n\rangle  \langle \nu_m| \hat P_{\Sigma}  \,e^{-{\beta \hat H}}\right), \label{eq:obs}
\end{align}
corresponding to a QMC observable  
$|\nu_n\rangle  \langle \nu_m| \hat P_{\Sigma}$ which is
a long chain of creation-annihilation operators.  In order to deal with such complicated observables, we developed an automated Wick decomposition which converts the observable in Eq.~\eqref{eq:obs} into a product of fermionic Green's functions with appropriate indices, which is then averaged over statistically generated auxiliary\khca{-}field configurations in \khcax{a}\khca{the} usual QMC process. The algorithm consists of the following stages: 1) conversion of the observable into a sum of products of creation and annihilation operators, each product arranged in inverse normal order, where all creation operators \khcax{are following}\khca{follow} all annihilation operators; 2) simplification of the expression removing all terms which compensate each other and joining together all products with the same operator structure; 3) applying a Wick decomposition to the resulting operator expressions. The algorithm is written using computer algebra of creation-annihilation operators, realized directly inside the QMC code \cite{superlat_github}. The only input needed for the computation \khcax{are}\khca{is} the algebraic expressions for the projector Eq.~\eqref{eq:proj_fin} and vectors Eq.~\eqref{eq:vec_fin} directly written in C++ code using overloaded arithmetic operators for objects of classes realizing linear operators acting on vectors in Fock space. The automated Wick decomposition for the arbitrary operator expression is explained in more \khcax{details}\khca{detail} below. 

Calculation of the trace in Eq.~\eqref{eq:obs} is done in QMC via a Trotter decomposition and subsequent Hubbard-Stratonovich (HS) decomposition of interaction terms:
\begin{align}
&\operatorname{Tr}\!\left(\hat O e^{-\beta \hat H}\right)
= \operatorname{Tr}\!\left(
\hat O \prod_{\tau}
\left[e^{-\Delta\tau\hat H_{tb.}}e^{-\Delta\tau H_{\mathrm{int}}}\right]
\right) \label{eq:trace}
\\
&\sim \sum_{\{e_x^\tau\}}
\prod_{\tau,x}\gamma\!\left(e_x^\tau\right)
\operatorname{Tr}\!\left(
\hat O \prod_{\tau} e^{-\Delta\tau\hat H_{tb}}
 e^{\sqrt{\Delta\tau\,U/2}\,\sum_x i\,\delta(e_x^\tau)\hat q_x}
\right),
\notag
\end{align}
where $\hat H_{tb.}$ is tight-binding part of the Hamiltonian \ref{eq:basic_ham}, $\Delta \tau$ is the step in the    and discrete charge-coupled auxiliary fields are used in the HS decoupling:
\begin{align}
\hat H_{\mathrm{int}} &= \frac{U}{2}\sum_x \hat q_x^{\,2},
\notag\\
\hat q_x &= \hat n_{x\uparrow}+\hat n_{x\downarrow}-1,
\end{align}
with coefficients $\gamma$ and $\delta$ defined as
\begin{align}
\gamma(\pm 1) &= 1+\frac{\sqrt{6}}{3},
&\gamma(\pm 2) &= 1-\frac{\sqrt{6}}{3},
\notag\\
\delta(\pm 1) &= \pm\sqrt{2(3-\sqrt{6})},
&\delta(\pm 2) &= \pm\sqrt{2(3+\sqrt{6})}.
\notag
\end{align}

The remaining trace is dealt with using Grassmann coherent states $|\chi_x\rangle$: a projector
\begin{align}
\hat I
&= \int \mathcal D\chi\,\mathcal D\bar\chi\,
 e^{-\sum_x \bar\chi_x\chi_x}
 |\chi_x\rangle\langle\bar\chi_x|\khcax{.}\khca{}
\notag
\end{align}
is inserted between each operator exponent in Eq.~\eqref{eq:trace}.

The observable $\hat O$ is represented as a sum of products with inverse normal order of creation and annihilation operators. Thus the projector can be inserted in the middle of the product:
\begin{align}
\hat c_{x_1}\hat c_{x_2}\cdots \hat c_{x_k}
|\chi_x\rangle\langle\bar\chi_x|
 \hat c_{y_1}^{\dagger}\cdots \hat c_{y_k}^{\dagger} \\
=\chi_{x_1}\cdots\chi_{x_k}
|\chi_x\rangle\langle\bar\chi_x|
\bar\chi_{y_1}\cdots\bar\chi_{y_k}.
\notag
\end{align}
Subsequently, the integral over Grassmann variables is taken using the expression:
\begin{align}
&\int \mathcal D\chi\,\mathcal D\bar\chi\,
\exp\!\left(
-\sum_{\substack{x,y\\ \tau_1,\tau_2}}
\bar\chi_x^{\tau_1}
\left[G^{-1}\right]_{xy}^{\tau_1\tau_2}
\chi_y^{\tau_2}
\right)
\chi_{x_1}^{\tau}\bar\chi_{y_1}^{\tau}\cdots
\chi_{x_k}^{\tau}\bar\chi_{y_k}^{\tau}
\notag\\
&\qquad=
\det G^{-1}\,
\det\!\begin{pmatrix}
G_{x_1y_1}^{\tau\tau} & \cdots & G_{x_1y_k}^{\tau\tau}\\
\vdots & \ddots & \vdots\\
G_{x_ky_1}^{\tau\tau} & \cdots & G_{x_ky_k}^{\tau\tau}
\end{pmatrix}, \label{eq:g_det}
\end{align}
where $G$ is the fermionic propagator for a given set of auxiliary fields.
As a result, each operator product in the observable $\hat O$ is converted into a determinant constructed from corresponding elements of the Green's function $G$. This conversion \fbc{has to be performed} only once at the start of the QMC simulation, for all operator expressions \fbc{of the} observables, and in practice, we only need to store \khcax{indexes}\khca{indices} of Green's functions in these determinants. 

For a 4-site reduced density matrix, we get $256^2 = 65536$ operators to compute. However, in practice, the majority of them \khcax{is}\khca{are} zero due to \khcax{the spin and particle number}\khca{spin and particle-number} conservation. These symmetries are automatically taken into account during conversion of operator expressions into a set of determinants Eq.~\eqref{eq:g_det}, and the total number of operators to compute is reduced to 2578. An additional technical trick is that for each set of parameters, two QMC simulations are run: one with spin\khca{-}coupled, and another one with charge\khca{-}coupled auxiliary fields. For some $\hat A_i$ operators, fluctuations can be drastically (up to two orders of magnitude) different depending on the HS channel used. For this reason, we perform postprocessing, where the final value of observable $\langle \hat A_i \rangle$ is automatically selected from simulations in the channel with smaller fluctuations for this specific $\hat A_i$.

In the present paper, we measure four-site density matrices
from a simulation of a system with altogether $16\times 16$ 
sites for the \emph{SM}-\emph{MI} and \emph{MI}-\emph{BI} transitions. Since we set \khca{the }interaction to 
zero \khcax{in the case of}\khca{for the} \emph{SM}-\emph{BI} transition, no actual QMC simulation
is done in that case. We simply substitute free fermionic 
propagators in the same automated Wick decomposition  Eq.~\eqref{eq:g_det}. In this case we show data for a $48\times 48$ system for the 4-site density matrix and for much larger systems (up to $1024 \times 1024$) for the entanglement entropy between two halves of the system (see below). The inverse temperature is set to $\beta=16$. We have  checked that lowering the temperature to $\beta=32$ does not 
noticeably change the results.
The two-site density matrices are then obtained by 
tracing out two of the lattice sites while respecting the
fermionic signs which are introduced in this process. Finally, we obtain error bars for the entanglement measures by bootstrapping over multiple independent QMC runs.

\section{Results} \label{sec:results}
\begin{figure*}
    \centering 
    \includegraphics[width=\linewidth]{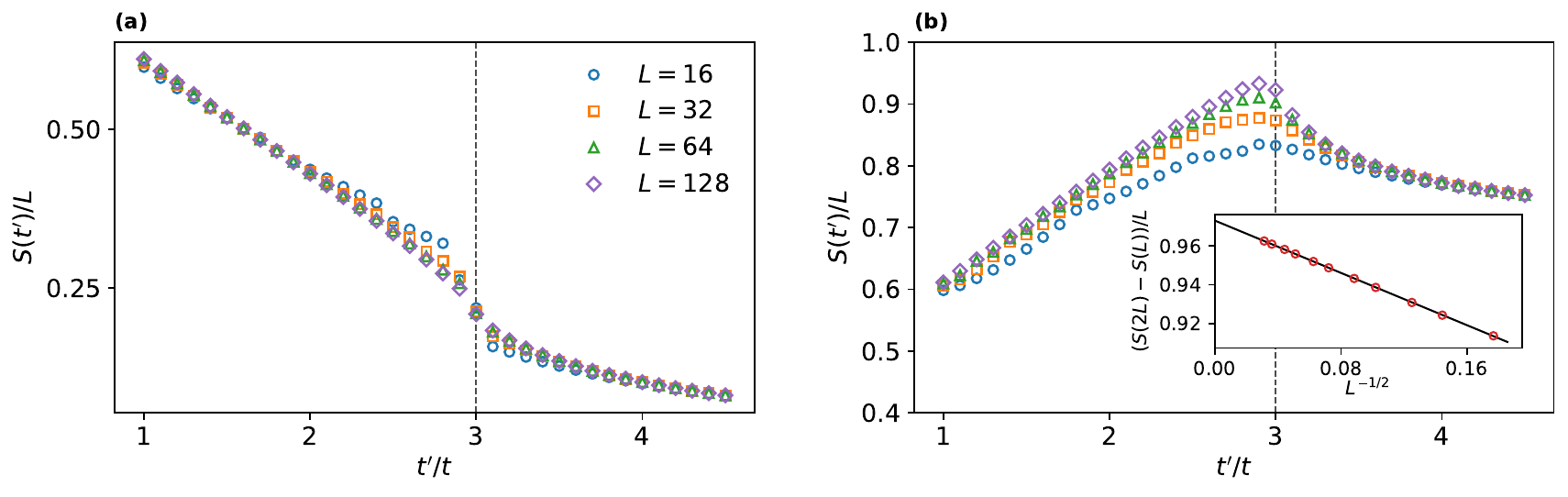}
    \caption{Entanglement entropy between  the two  halves of the system. 
    Panel (a): Entropy according to a cut in the $x$\kh{-}direction where 
    the subsystem is given by $y<L/2$. 
    Panel (b): Entropy according to a cut in the $y$\kh{-}direction where 
    the subsystem is given by $x<L/2$.  The inset shows the size
    dependence of the subleading term, at the $t'=3$ \emph{SM} to \emph{BI} (dashed line) transition using system sizes up to $L=1024$}
    \label{fig:ent_entropy}
\end{figure*}
\begin{figure*}
    \centering
        \includegraphics[width=\linewidth]{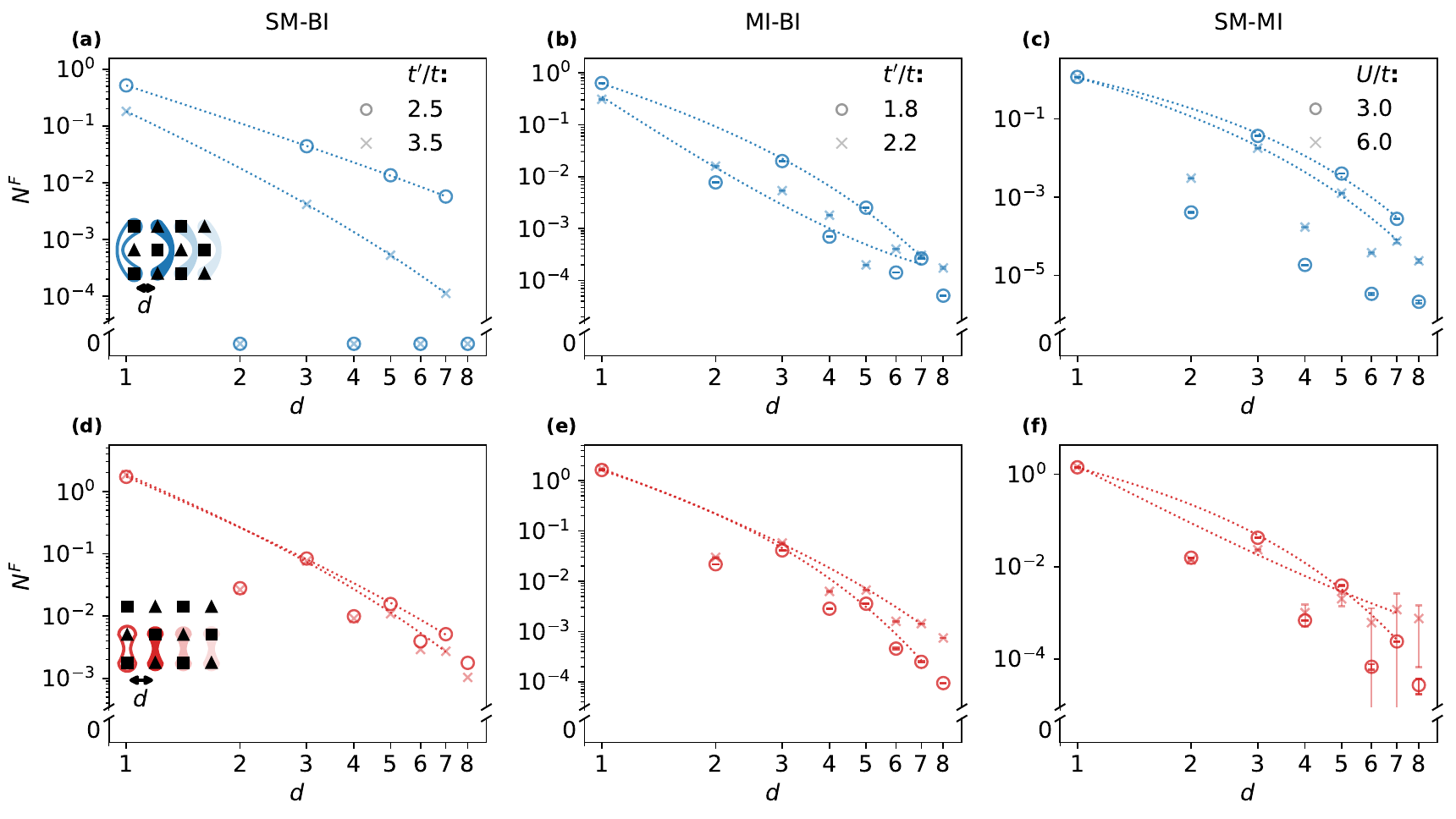}
        \caption{
        Fermionic negativity $N^\mathrm{F}$ between dimer-dimer configurations
        \kh{as a function of distance $d$} for the three transition types \emph{SM}-\emph{BI} \kh{(left)}; \emph{MI}-\emph{BI} \kh{(center)}; \kh{and}  \emph{SM}-\emph{MI} \kh{(right)} before and after the transition. \kh{Top:} each dimer only spans lattice sites of the same sublattice\kh{; bottom:} each dimer couples both sublattices. Values smaller \kh{than} $10^{-8}$ are indicated as zero. Inset\kh{s}: We measure entanglement between the hollow highlighted sites \kh{(subsystem $A$)} and the filled highlighted sites \kh{(subsystem $B$)}. Fading indicates  \kh{increasing distances of subsystem $B$}. Squares are \kh{$\cal A$-sublattice sites} while triangles are \kh{$\cal B$-sublattice sites}.
        Dotted lines display a $N^\mathrm{F}(d) = \kappa\, d^{-\epsilon} e^{-d/\xi}$ fit to the data with parameters given in Tab.~\ref{tab:fits_SM_BI}-Tab.~\ref{tab:fits_SM_MI}, evidencing the transition away from a pure power law behavior of the \emph{SM} phase.}
        \label{fig:dAAAA}
\end{figure*}
\begin{figure*}
    \centering
        \includegraphics[width=\linewidth]{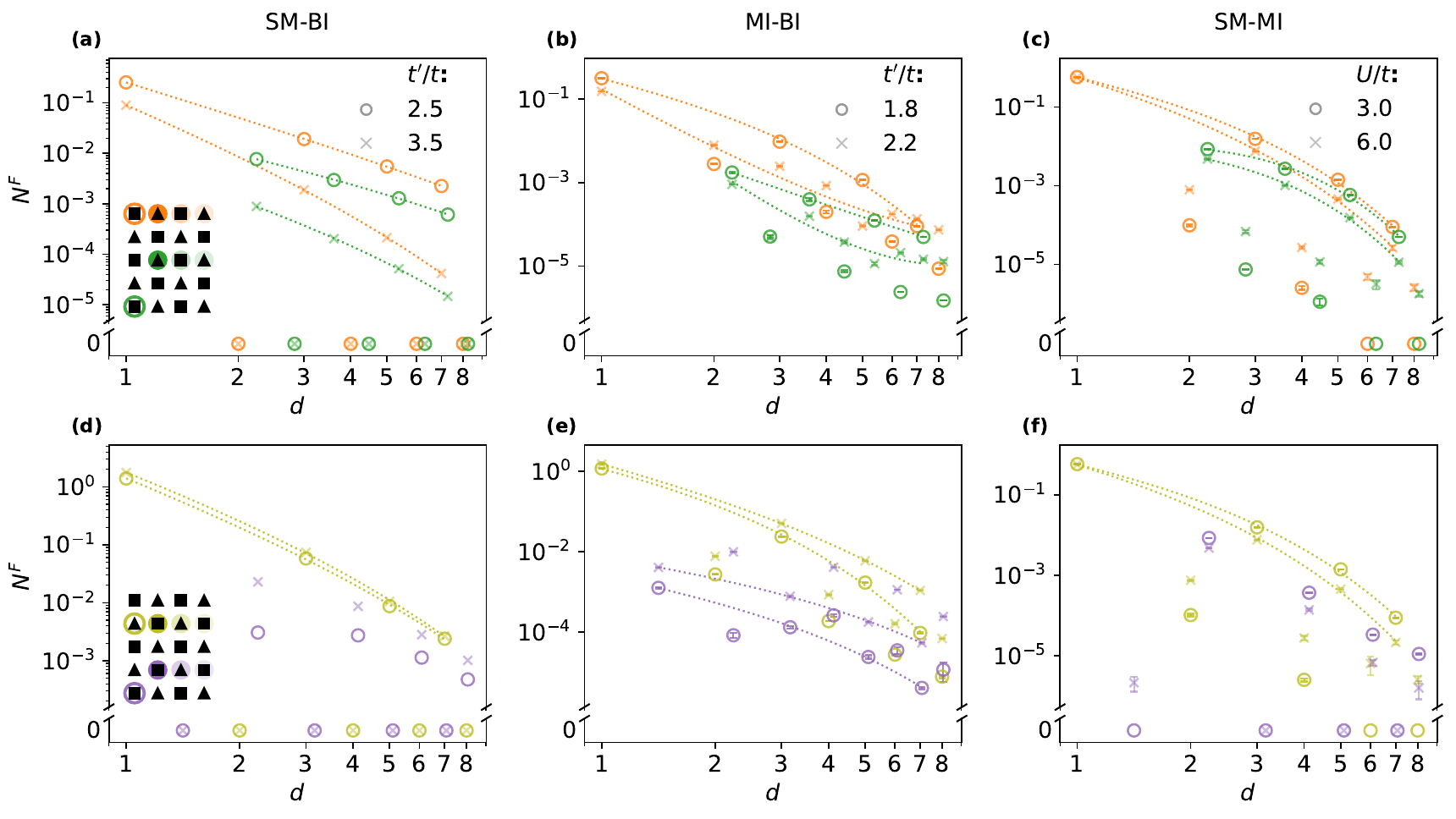}
        \caption{Same as Fig.~\ref{fig:dAAAA} but for the 2s configuration\kh{s (see inset).
        \kh{Top:} sites are \kh{separated} 
        \kh{with}  negative hopping
        parameters alternating  \kh{with distance} between $-t,-t'$. \kh{Bottom:} sites are 
        \kh{separated} \kh{with} positive hoppings alternating
        between $t,t'$} 
        }
        \label{fig:dABAB}
\end{figure*}

\begin{figure}
    \centering
    \includegraphics[width=\linewidth]{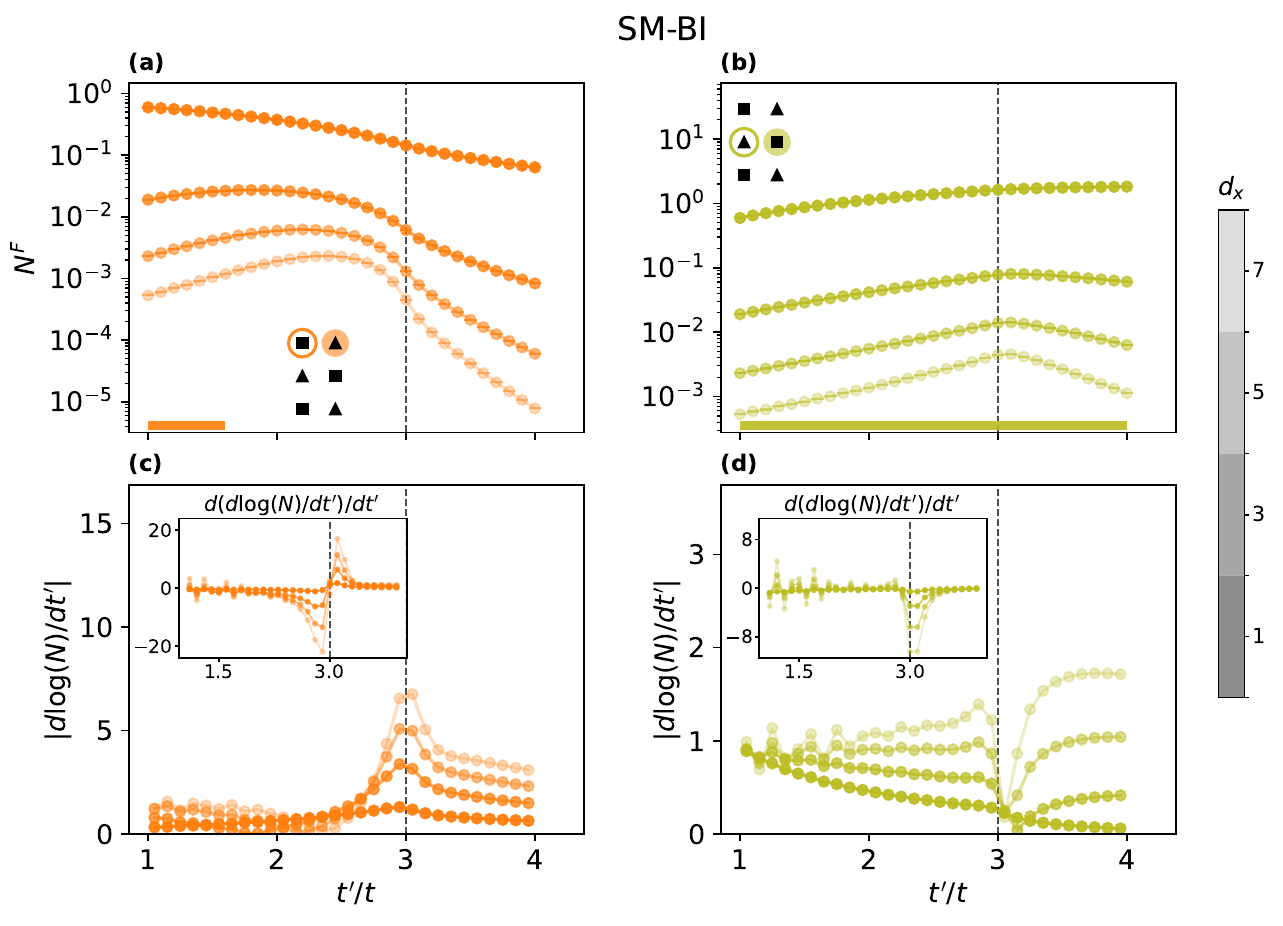}
    \caption{(a,b) 2s $N^\mathrm{F}(t')$ for the \emph{SM}-\emph{BI} transition for different non-zero distances \kh{(according to the color intensity bar)}; (c,d) $\partial\log(N^\mathrm{F})/\partial t'$ with the inset showing
    $\partial^2\log(N^\mathrm{F})/\partial t'^2$. Colors encode the geometric configuration 
    of the two sites (see insets \fbc{here and} in Figs.~\ref{fig:dABAB},~\ref{fig:dAAAA} (d)). The 
    horizontal bar along the x-axis indicates the regime of a non-zero spin entanglement $N^\mathrm{S}$.  \fbc{And we indicate by the vertical line the true transition as identified in Appendix~\ref{Appendix_phase_transitions}}.
    }
    \label{fig:2s_NF_SM_BI}
\end{figure}

\begin{figure}
    \centering
    \includegraphics[width=\linewidth]{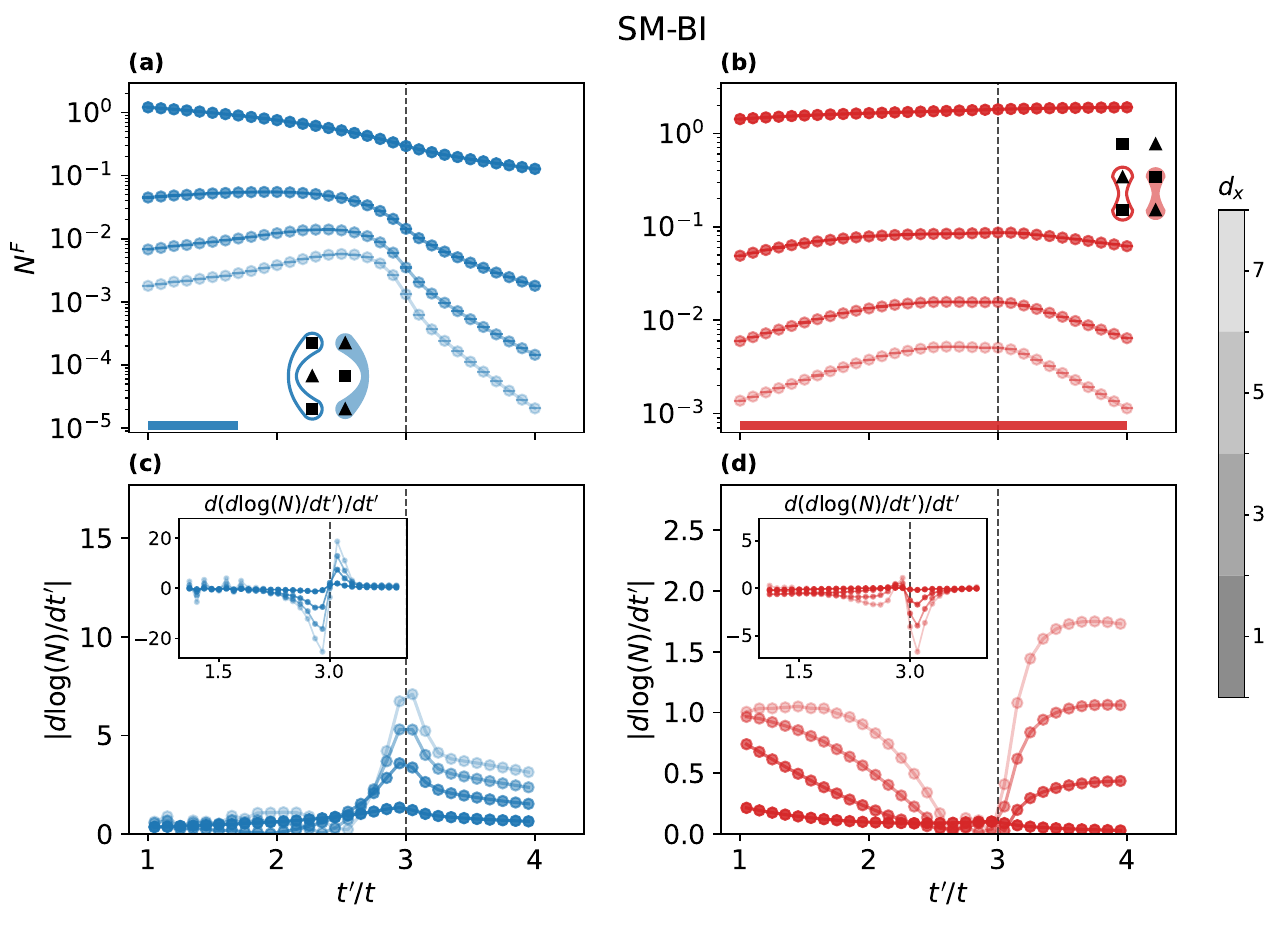}
    \caption{Same as Fig.~\ref{fig:2s_NF_SM_BI} but for dimer-dimer 
    entanglement 
    across the \emph{SM}-\emph{BI} transition.
    }
    \label{fig:dimer_dimer_NF_SM_BI}
\end{figure}

\subsection{SM-BI transition}
\subsubsection{Entanglement entropy}
The simplest of the three transitions we consider in this paper 
is the \emph{SM-BI} transition\khcax{, nevertheless}\khca{; nevertheless,} understanding it in detail 
at $T=0$ provides a controlled reference point before turning to 
the finite-temperature case, where the negativity-based measures
introduced above become necessary. We track the transition in the 
non-interacting limit at $U=0$. In \kh{contrast} to the other 
transitions, we can solve it exactly and therefore obtain the 
ground\khca{-}state entanglement entropy. Since the entanglement entropy
is a faithful entanglement measure only for pure states, we evaluate
it exclusively in the non-interacting limit at zero temperature, 
where the whole state is pure.  Being able to calculate the von 
Neumann entanglement entropy in the $t'$-$\pi$-flux model has the 
advantage that we have a more conventional entropy as a reference point,
before turning to the dimer-dimer and 2s entanglement measures based on the negativity. 

In Fig. \ref{fig:ent_entropy}\kh{,} the entanglement entropy  is 
shown for various system sizes and $t'$ parameters. While for 
systems with a finite Fermi surface the entanglement entropy is 
known to have a $S \sim L^{d-1} \log L$ scaling and for gapped 
systems a strict boundary law $S \sim L$ \cite{EE_area_law_violation_2006_I,EE_area_law_violation_2006_II,EE_area_law_violation_2007_III}, 
the case of discrete Dirac points is between the two regimes.
Our numerical results support the
\begin{equation}
S \sim a L
\end{equation}
area law for the leading order\kh{,} both in the semimetal and 
the band\kh{-}insulator cases.
If the boundary of the two subsystems is in the $x$-direction
\kh{[}see panel (a) of Fig. \ref{fig:ent_entropy}\kh{]}, the
derivative of the entanglement entropy with respect to  \kh{$t'$} has a peak at the transition point \khca{for} small lattices, but subsequently it becomes smoother and smoother so the initial peak gradually disappears in the thermodynamic limit. On the contrary, if the boundary of the two subsystems is in the \kh{$y$-}direction, the entanglement entropy \kh{instead} shows a maximum at the $t'\!=\!3$\kh{-}transition, which seemingly develops into a discontinuity in the thermodynamic limit.  The difference between cuts in \khcax{X- and Y- directions}\khca{the $x$- and $y$-directions} is connected to the asymmetry of the $t'$-$\pi$\khcax{-model}\khca{-flux model}: once \khca{the }Dirac cones approach each other and collide, the dispersion in the \khcax{Y}\khca{$y$} direction (direction of collision) becomes quadratic, while the dispersion in the perpendicular direction remains linear.

The question of the subleading correction is more \kh{subtle}.
At the transition between the semimetal\kh{lic} and band\kh{-}insulat\kh{ing} phase, where the two Dirac cones merge,
we find the following scaling if one cuts the system along the \kh{$y$-}direction\kh{:}
\begin{equation}
S = a L + b L^{0.5 \pm 0.02}.
\label{eq}
\end{equation}
This is demonstrated in the inset of panel (b) of Fig.~ \ref{fig:ent_entropy}. 
Here\kh{,} we plot \khcax{the difference of}\khca{} $(S(2L) - S(L))/L$ as a function of the 
inverse square root of the system size $1/\sqrt{L}$. The difference \kh{takes} 
the leading order out, such that we can investigate the subleading term.
A \khcax{precise}\khca{straight} line is obtained, implying the scaling given in Eq.~\eqref{eq}.

In the band\kh{-}insulator phase we find\kh{,} in both directions,  the subleading order to be a constant $O(1)$ number. In the semimetal phase, if we cut in the 
\kh{$x$-}
direction, the subleading term is a constant; if we cut in the $y$-direction, the 
subleading term is oscillating periodically with $L$, with $O(1)$ amplitude. Our 
findings are compatible with the entanglement entropy results in Dirac-cone systems 
\cite{EE_Dirac_phase_trans_2024, EE_Dirac_2012, EE_Dirac_PRL_2008}, which investigate slightly different models with cone-like spectra.

\subsubsection{Distance scaling of the fermionic negativity}
We now turn to the objective of this paper and consider finite temperatures $T=1/16\khca{\;}t$.
At finite temperature, the entanglement entropy is not a valid measure of entanglement 
any longer. We therefore now consider the fermionic negativity and restrict ourselves 
to small subsystems of either four or two lattice sites \kh{(later denoted as 
dimer-dimer or 2s entanglement) at distance $d$}. The specific geometric configurations we 
consider are indicated in the insets of Figs.~\ref{fig:dAAAA}~and~\ref{fig:dABAB}.
The 
color scheme introduced here will be carried through the paper encoding the specific spatial configurations: 

\begin{minipage}{8.2cm}
\begin{itemize}
    \item[\textcolor{fitblue}{\emph{ blue}}] for dimer-dimer entanglement where each dimer couples sites on the same sublattice (triangles or \khcax{circles}\khca{squares})
    \item[\textcolor{fitred}{\emph{ red}}] dimer-dimer configuration but each dimer mixes the sublattices
    \item[\textcolor{fitorange}{\emph{ orange}}] 2s entanglement with alternating sublattice on the second site which we move along the horizontal direction following a path of positive $t,\,t'$ hoppings
    \item[\textcolor{fitgreen}{\emph{ green}}] same as \textcolor{fitorange}{orange} but with the second site shifted by $2$ sites in the $y$-direction
    \item[\textcolor{fitkhaki}{\emph{khaki}}] same as \textcolor{fitorange}{orange} but following a path of negative $-t,\,-t'$ hoppings (see inset in Fig.~\ref{fig:phase_digram})
    \item[\textcolor{fitpurple}{\emph{ purple}}] same as \textcolor{fitkhaki}{khaki} with a $y$-direction shift of the second site by $1$ also following a path of negative $-t,\,-t'$ hoppings
\end{itemize}
\end{minipage}
Note that this only includes configurations where the true distance between
subsystems $A$ and $B$ increases with $d$. By this, we mean that no entanglement 
is mediated purely by the geometry, as it would be the case for example 
in a cross\-shaped configuration with one site of $A$ next to one of $B$ 
even at large distances.
In Figs.~\ref{fig:dAAAA} and~\ref{fig:dABAB} we show the distance dependence of these six geometric configurations of entanglement 
measured by the fermionic negativity for all three types of phase 
transition. We show one set of data before and one set of data 
after the transition\kh{, corresponding to the beginning and end of the red arrows in Fig.~\ref{fig:phase_digram}}. Note that the spin entanglement $N^\mathrm{S}$  (not shown) is always zero beyond the first distance \fbc{\footnote{In the 2D Heisenberg AFM two-site entanglement also only exists between nearest \khca{neighbors} \cite{Subrahmanyam_2004}}}. 
In a simplified view, the negativity is nothing more than a 
resummation of correlation functions\kh{, though a specific one,
that allows us to witness entanglement between subsystems}. Hence
any universal behavior of the correlation functions is expected to carry over to it. 
For $U=0$ this statement is exact and can be proven for the 2s case at half filling:
Let $i\in A$ and $j\in B$ denote the two sites and $G_{ij}=\langle \hat c^\dagger_{i\sigma}\hat c_{j\sigma}\rangle$ the equal-time Green's function between them, the simplest of the correlation functions ${\cal C}$ discussed below. Besides $G_{ij}$ and the occupations, $\rho_{AB}$ also contains two-particle matrix elements, e.g., spin-flip terms $\langle \hat c^\dagger_{i\uparrow}\hat c_{i\downarrow}\hat c^\dagger_{j\downarrow}\hat c_{j\uparrow}\rangle$ and pair-hopping terms $\langle \hat c^\dagger_{i\uparrow}\hat c^\dagger_{i\downarrow}\hat c_{j\downarrow}\hat c_{j\uparrow}\rangle$. At $U=0$, however, all these elements factorize by Wick's theorem into products of $G_{ij}$ and \fbc{SU(2) symmetry \khca{and half filling} provide occupation numbers} $\langle \hat n_{i\sigma}\rangle=\langle \hat n_{j\sigma}\rangle=1/2$. Therefore, $\rho_{AB}$ factorizes into independent spin-up and spin-down parts. If $G_{ij}=0$, the correlation matrix is block diagonal \fbc{and} $\rho_{AB}=\rho_A\otimes\rho_B$ is a product state \fbc{without entanglement.} For $G_{ij}\neq0$, the partial time reversal of each spin part has trace norm $1+4|G_{ij}|^2$, so that  $\|\rho_{AB}^{\tau_B}\|_{\rm tr}=(1+4|G_{ij}|^2)^2$ and 
\begin{equation}\label{eq_NF_from_G}
    N^\mathrm{F} = 2\log_2\!\left(1+4|G_{ij}|^2\right) \simeq \frac{8}{\log 2}\,|G_{ij}|^2,
\end{equation}
where the last step holds for $|G_{ij}|\ll 1$
\cite{Matsuda2025negativity,Shapourian_2019_finiteT}. Hence, at $U=0$ any non-zero $G_{ij}$ entails entanglement, and $N^\mathrm{F}$ decays with twice the exponent (or decay rate) of $G_{ij}$. Particle-hole symmetry further implies $G_{ij}=0$ for two sites on the same sublattice, which explains the vanishing 2s $N^\mathrm{F}$ for these configurations. For $U>0$, the connected parts of the two-particle elements are independent of $G_{ij}$, and $N^\mathrm{F}$ is no longer a function of $G_{ij}$ alone (see the \emph{MI}-\emph{BI} discussion below).

In the \emph{SM} phase,  we expect  a polynomial decay of correlation functions ${\cal C}$ with distance $d$,  specifically 
\begin{align}
{\cal C}(d) \sim d^{-2}
\label{eq:G_polynom}
\end{align}
for the 2D Dirac semimetal  \cite{PhysRevB.99.125145}. On the other hand, the opening of a band gap leads to an exponential decay with distance\khcax{,}\khca{;} fermionic correlations behave as 
\begin{align}
{\cal C}(d) \sim d ^{-\alpha} e^{-d/\xi(\Delta, m_x, m_y)}
\label{eq:G_polynom_exp}
\end{align}
in the \emph{BI} \kh{phase}, where $m_x$ and $m_y$  are effective masses of quadratic dispersion in two directions, $\Delta$ is a gap and \fbc{the} correlation length $\xi$ \fbc{is} some function which depends on the details of the model as well as \khca{the }shape and size of the subsystems $A$ and $B$. Here, $\alpha$ is the degree of polynomial correction, which is also system-dependent. 
In both cases $d$ and the corresponding correlation length $\xi$ as well as the degree of polynomial correction $\alpha$ correspond \fbc{ only to a separation between subsystems along the $x$-direction}. Another direction would have different constants due to the violation of rotational symmetry by the colliding Dirac cones. 

The asymptotic behavior of Eq.~\eqref{eq:G_polynom_exp} is clearly satisfied by the $N^\mathrm{F}$ data across the \emph{SM-BI} transition. The dimer-dimer case with dimers spanning the same sublattice (blue color) is shown in Fig.~\khca{\ref{fig:dAAAA}} (a): The dotted line indicates
a fit to $N^\mathrm{F}(d) = \kappa\, d^{-\epsilon} \fbc{e^{-d/\xi}}$,  which is essentially driven by a power-law behavior before the transition
at $t'=2.5$ \fbc{(dots)} but deviates from it in the \emph{BI} regime $t'=3.5$ \fbc{(crosses)}, now with a relevant exponential correction \fbc{$e^{-d/\xi}$} (see Tab.~\ref{tab:fits_SM_BI}). The same behavior is also seen in the 2s data in Fig.~\ref{fig:dABAB} (a) (orange) \fbc{and to a lesser \khcax{extend}\khca{extent} in the \khcax{dimer mixing (red) configuration}\khca{mixed-sublattice dimer (red) configuration} as well as the green and khaki 2s configurations (see Tab.~\ref{tab:fits_SM_BI})}. 

\begin{table}[t]
\centering
\caption{Fits of $N^\mathrm{F}(d) = \kappa\, d^{-\epsilon} \fbc{e^{-d/\xi}}$ for the \emph{SM}-\emph{BI} transition.}
\label{tab:fits_SM_BI}
\begin{tabular}{l c | c c}
partition & $t'/t$ & $\epsilon$ & \fbc{$\xi$}  \\
\hline
\hline
\textcolor{fitblue}{blue}     & 2.5 & 2.12 & 15.87 \\
                              & 3.5 & 2.91 & 3.51 \\
\hline
\textcolor{fitorange}{orange} & 2.5 & 2.23 & 15.87 \\
                              & 3.5 & 2.92 & 3.0 \\
\hline
\textcolor{fitgreen}{green}   & 2.5 & 1.59 & 7.87 \\
                              & 3.5 & 2.14 & 3.24 \\
\hline
\textcolor{fitred}{red}       & 2.5 & 2.47 & 5.75 \\
                              & 3.5 & 2.33 & 2.97 \\
\hline
\textcolor{fitkhaki}{khaki}   & 2.5 & 2.43 & 3.58 \\
                              & 3.5 & 2.31 & 2.95 \\
\end{tabular}
\end{table}

We should also note that there is \fbc{an additional} power law exactly at criticality, since when the  two Dirac cones collide,  the  density of \khca{states} scales as $N(\omega) \propto \sqrt{\omega}$ \cite{Ixert_piflux_2014}. \khca{As mentioned above,} exactly at criticality, the dispersion  becomes quadratic in \khca{the }\khca{$y$} direction, but remains linear in \khca{the }\khca{$x$} direction, which \khca{also} implies \khca{}different power laws for correlation functions in different directions. However, as we will \khca{show} below, the \khca{exponential correction} dominates the changes in power law at large distances.  

The switch from polynomial Eq.~\eqref{eq:G_polynom} to combined exponential-polynomial  behavior Eq.~\eqref{eq:G_polynom_exp} can be demonstrated more explicitly, by plotting the fermionic negativity \khcax{depending on t'}\khca{as a function of $t'$} across the \emph{SM}-\emph{BI} transition.
Figs.~\ref{fig:2s_NF_SM_BI}~and~\ref{fig:dimer_dimer_NF_SM_BI} (a) and (b) show this evolution of 2s $N^\mathrm{F}$ with $t'/t$ for various distances $d_x$.
Assuming the general form
\begin{align}
 N^\mathrm{F}\kh{=}\kappa\, \frac{1}{d^\epsilon} e^{-d/\xi},
\end{align}
we can compute the logarithmic derivative of the fermionic negativity \khcax{over t'}\khca{with respect to $t'$} as
\begin{align}
 \frac{\partial \log N^\mathrm{F}}{\partial t'} = \frac{\partial \log \kappa}{\partial t'} - \frac{\partial \epsilon}{\partial t'} \log d + \frac{d}{\xi^2} \frac{\partial \xi}{\partial t'} .   \label{eq:log_der}
\end{align}
For large distances, the last term, which stems from the exponential part of the asymptotics, dominates, since $d\gg \log d$ in this limit. 
The correlation length $\xi$ diverges once we approach the phase transition from the  \emph{BI} side. Since the gap  $\Delta$ is linearly dependent on $t'$, the simplest form of the correlation length would be $\xi=(t'-3)^{-\nu}$ with $\nu=1$.  \khcax{It leads to the fermionic negativity}\khca{This gives}
\begin{align}
 \frac{\partial \log N^\mathrm{F}}{\partial t'} \sim d\, (t'-3)^{\nu-1} = d .   \label{eq:log_der_fin}
\end{align}
 in the large $d$ limit in the vicinity of the phase transition \khcax{at the  \emph{BI} side of it}\khca{on its \emph{BI} side}. On the other hand, the exponential term is \gbc{negligible} on the \emph{SM} side of the transition, \gbc{at the accessible distances}, hence the logarithmic derivative switches from logarithmic (due to \khcax{sub leading}\khca{subleading} terms in  Eq.~\eqref{eq:log_der}) to linear dependence on $d$ at the transition point. It leads to a peak growing linearly with distance in the derivative of the logarithmic derivative, right at the \emph{SM}-\emph{BI} transition. \khcax{And this}\khca{This} is exactly the behavior we see in the insets in \khcax{Fig.~}\khca{Figs.~}\ref{fig:2s_NF_SM_BI}(d) and  \ref{fig:dimer_dimer_NF_SM_BI}(d): they both show the peaks in the second derivative located exactly at the transition point at $t'=3$ with height of the peaks increasing roughly linearly with distance $d$.  Hence for these two shapes of the $A$ and $B$ subsystems, with the separation in this particular direction, $\nu$ is indeed 1. We should note, that due to already mentioned non-equivalence of $x$ and $y$ direction, $\nu$ value can vary for different separation directions. 
 
 The other two shapes (Fig.~\ref{fig:2s_NF_SM_BI}(c) and \ref{fig:dimer_dimer_NF_SM_BI}(c)) again show a switch to a linear dependence of the logarithmic derivative on $d$ at the transition. For smaller $t'<3$, the logarithmic derivative does not exhibit a clear dependence on the separation $d$, while after the transition, for $t'>3$, the $d=1,3,5,7$ lines are roughly equidistant, signaling the appearance of the term $\frac{d}{\xi^2} \frac{\partial \xi}{\partial t'}$. However, they also show an additional peak in the logarithmic derivative itself right at the phase transition, suggesting a more complicated dependence of the correlation length on the effective masses and the gap in this case.

As hypothesized in the introduction, the transition is visible already in the smallest 2s case 
(Fig.~\ref{fig:2s_NF_SM_BI}) and does not require an investigation of the system-size 
scaling, rather a distance scaling is enough to unveil the transition. \fbc{In \khcax{Appendix.~}\khca{Appendix~}\ref{Appendix_I} we show that the same signatures are also picked up by the mutual information, a measure of all correlations between \khcax{two bipartitions}\khca{the two subsystems}.}

Before discussing the remaining transitions, lets first investigate the observed physics in more detail and compare to the \emph{plain-vanilla} 2D Hubbard model.
In the non-interacting case $U=0$,
the sublattice structure of the problem has a peculiar consequence\kh{:} entanglement only 
exists between two lattice sites with an antiferromagnetic correlation between them 
(see Fig.~\ref{fig:dABAB}(a),(d)), i.e. sites on different sublattices\kh{. T}his is
due to the antiferromagnetic tendency of the model which was also observed to lead
to the same structure in the \emph{plain\kh{-vanilla}} 2D Hubbard model 
\cite{Bippus_2sRDM_entanglement_2026}\fbc{. The formal reason is given by Eq.~\eqref{eq_NF_from_G}, since in} the free case on a bipartite lattice, the fermionic Green's function is exactly zero for sites located on the same sublattice. Consequently, \fbc{there is also no entanglement }
between dimers that are fully located on the same sublattice (see Fig.~\ref{fig:dAAAA} 
(a)), while the mixed sublattice structure for the geometry explored in panel (d) ensures 
that the entanglement always remains finite. Curiously, in contrast to the 
\emph{plain\kh{-vanilla}} 2D Hubbard model where increased interactions do not 
reinstate entanglement for sites with ferromagnetic correlations 
\cite{bippus2026_qsim}, we here observe such \khcax{an}\khca{a} behavior. 

Similarly, the $\pi$-flux model differs from the \emph{plain\kh{-vanilla}} 
2D Hubbard model by the \kh{existence of 2s spin entanglement $N^{\mathrm S}$ between
nearest \khca{neighbors} in the \kh{small}-$U$ region} 
\cite{Bellomia2023,BellomiaPhD,Bippus_2sRDM_entanglement_2026}, as
indicated by the horizontal bar in Fig.~\ref{fig:2s_NF_SM_BI} (a) and (b).  
 Note that any spin entanglement in the $t'$-$\pi$-flux model dies 
beyond the nearest-\khca{neighbor} distance (therefore we do not show it here), and any fermionic entanglement beyond
this distance is purely mediated by hopping of the electrons. 
Once a $t'$ is turned on with a sufficient strength, the spin entanglement dies for configurations where the two subsystems $A$ and $B$ are connected by positive hopping amplitudes $t,t'$.
This is due to the following effect\khca{:} an 
increased hopping $t'$ leads to a strong antiferromagnetic exchange coupling 
along its bond, beating the coupling along the $t$ bond. Since the spin 
entanglement of the \emph{orange} nearest-\khca{neighbor} configuration sits on such a $t$ 
bond, it vanishes with increased $t'$. Specifically, this happens when spin-triplet 
correlations dominate over the spin singlet \cite{bippus2026_qsim}.
\khcax{Note-worthily}\khca{Notably}, the sudden death of \kh{spin} entanglement happens before the phase 
transition\khca{,} which further demonstrates the usefulness of the fermionic negativity to 
capture the full physics of the considered model, albeit not providing any operationally
useful resource \cite{Ding2022QuantifyingEE}.
\khcax{Vice-versa}\khca{Conversely}, the nearest-\khca{neighbor} 2s spin entanglement along the $|t'|$ bond 
survives across the transition (\emph{khaki} configuration).
The death of the spin dimer-dimer entanglement (see horizontal bar in 
Fig.~\ref{fig:2s_NF_SM_BI}) can then be \khcax{directly be explained through the behaviors}\khca{explained directly by the behavior} of the
involved 2s entanglement. 

\subsection{MI-BI transition}
\begin{figure}
    \centering
    \includegraphics[width=\linewidth]{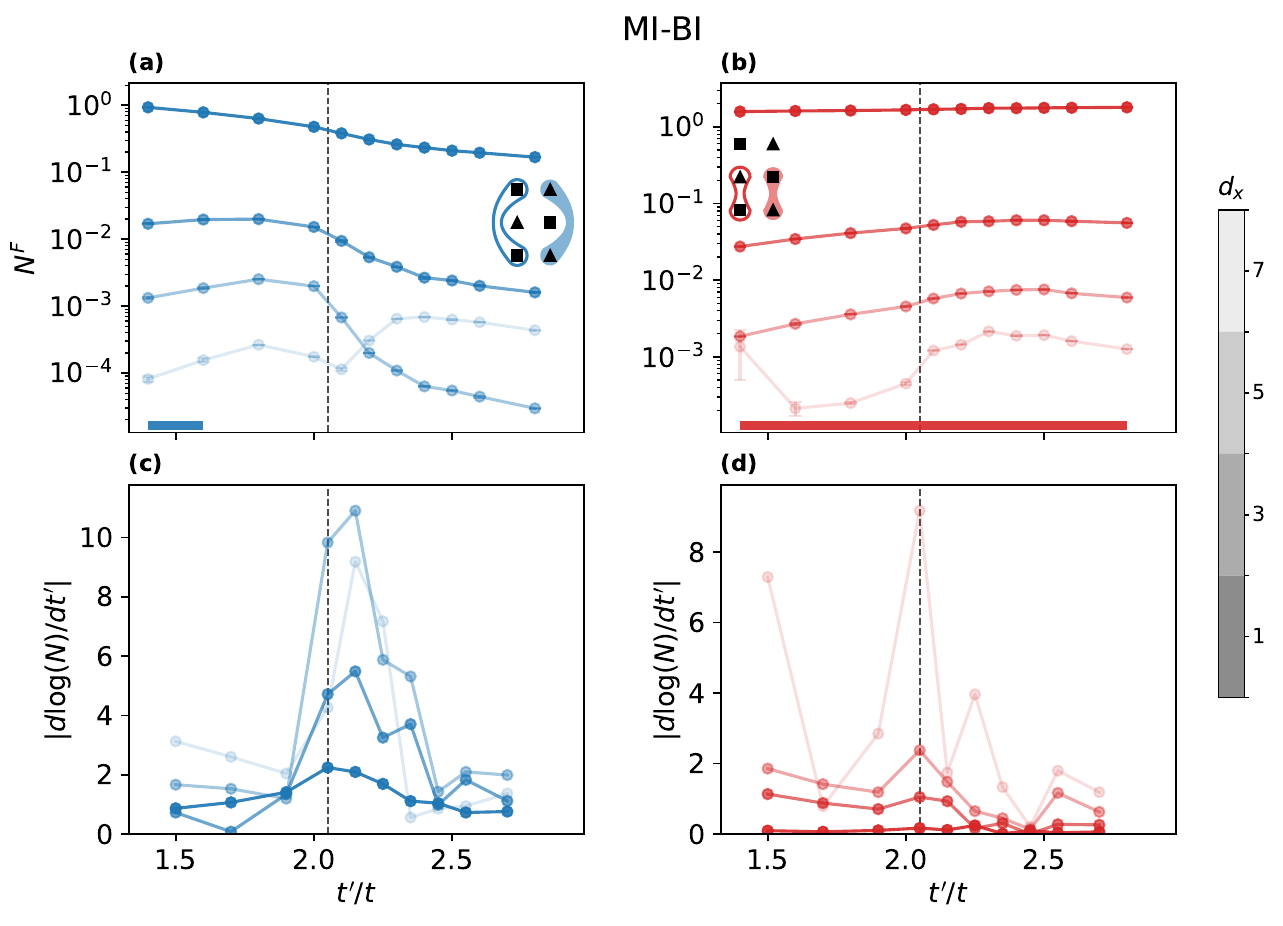}
    \caption{Same as Fig.~\ref{fig:2s_NF_SM_BI} and Fig.~\ref{fig:dimer_dimer_NF_SM_BI} but for dimer-dimer entanglement across the \emph{MI}-\emph{BI} transition.}
    \label{fig:dimer_dimer_NF_MI_BI}
\end{figure}

\begin{figure}
    \centering
    \includegraphics[width=\linewidth]{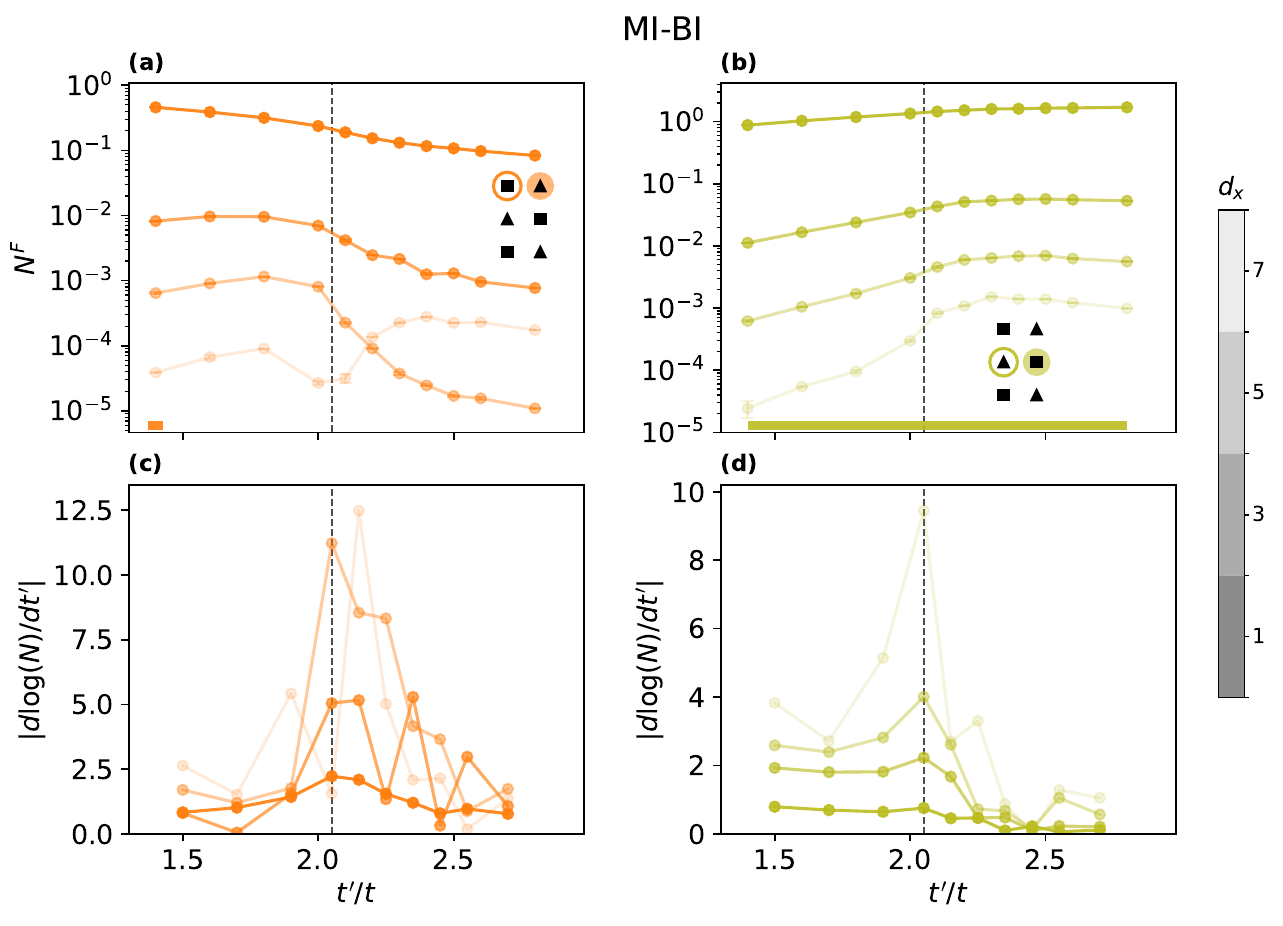}
    \caption{2s $N^\mathrm{F}$ and logarithmic derivatives thereof for the \emph{MI}-\emph{BI} transition.}
    \label{fig:2s_NF_NI_BI}
\end{figure}

\begin{table}[t]
\centering
\caption{Fits of $N^\mathrm{F}(d) = \kappa\, d^{-\epsilon} \fbc{e^{-d/\xi}}$ for the \emph{MI}-\emph{BI} transition.
$^\dagger$Unphysical fit ($\xi < 0$).}
\label{tab:fits_MI_BI}
\begin{tabular}{l c | c c}
partition & $t'/t$ & $\epsilon$ & {$\xi$}  \\
\hline
\hline
\textcolor{fitblue}{blue}     & 1.8 & 1.71 & 1.26 \\
                              & 2.2 & $5.16^\dagger$ & $-2.30^\dagger$ \\
\hline
\textcolor{fitorange}{orange} & 1.8 & 1.42 & 1.14 \\
                              & 2.2 & $5.23^\dagger$ & $-2.27^\dagger$ \\
\hline
\textcolor{fitgreen}{green}   & 1.8 & $3.17^\dagger$ & $-25^\dagger$ \\
                              & 2.2 & $7.84^\dagger$ & $-1.07^\dagger$ \\
\hline
\textcolor{fitred}{red}       & 1.8 & 1.46 & 1.00 \\
                              & 2.2 & 2.38 & 2.41 \\
\hline
\textcolor{fitkhaki}{khaki}   & 1.8 & 1.46 & 0.92 \\
                              & 2.2 & 2.21 & 2.03 \\
\hline
\textcolor{fitpurple}{purple} & 1.8 & 1.38 & 1.63 \\
                              & 2.2 & 1.14 & 2.25 \\
\end{tabular}
\end{table}

Let us now turn to the \emph{MI-BI} transition. We apply the same detection technique as in the previous section, computing the logarithmic derivative of the fermionic negativity $N^\mathrm{F}$ with respect to $t'$ across the transition. Calculations are performed at $U=7$, so the results are from full QMC simulations. Figure \ref{fig:dimer_dimer_NF_MI_BI} shows the logarithmic derivative of $N^\mathrm{F}$ for dimer-dimer entanglement, and {Fig.~}\ref{fig:2s_NF_NI_BI} shows the same for {2s} entanglement.
We again see a peak growing with the distance exactly at the transition point. We note that since we are more limited in system size in QMC simulations (calculations were done on a $16\times16$ lattice instead of the $48\times48$ one for \emph{SM-BI} transition), we expect more finite-size effects. In particular, the data at $d=7$ {are} less reliable, since this separation is  already comparable to half the lattice size.  Nevertheless, even the present data shows a peak growing with $d$ at the transition point, consistent with a divergence in the $d\rightarrow\infty$ limit.

For the interpretation of the data\khca{,} we \khcax{can not}\khca{cannot} rely on two-point correlation functions alone, since unlike the previously considered \emph{SM-BI} case this is an interacting system. For the \emph{MI}, while fermions have a gap, spin-spin correlations still have a polynomial form. In the \emph{BI} case, all correlation functions should acquire an exponential \khcax{asymptotic in the large \fbt{separation} limit}\khca{asymptotics at large separation}. However, due to the coupling of electrons to critical spin fluctuations, the correlation length in these exponents can behave non-analytically at the phase transition.

Let us consider how this change in correlations can be reflected in the behavior of $N^\mathrm{F}$. First, we note that the spin entanglement $N^\mathrm{S}$ still disappears for all distances beyond nearest\khcax{-} neighbors. The $N^\mathrm{S}$  negativity is obtained from $\rho^\mathrm{S}_{AB}$, which is the reduced density matrix after application of the selection rules Eq.~\eqref{eq:sel_rules}. These selection rules remove all \khcax{non-diagonal}\khca{off-diagonal} elements of the reduced density matrix except those proportional to spin-spin correlations between subsystems. \khcax{It also leads to hermiticity of the partially transposed density matrix which the negativity $N^\mathrm{S}$ is obtained from.}\khca{They also make the partially transposed density matrix, from which $N^\mathrm{S}$ is obtained, Hermitian.} This is essentially the mathematical origin of the sudden death, since for \khcax{hermitian matrix}\khca{a Hermitian matrix} the negativity is exactly zero if all eigenvalues are positive. This happens for all $d>1$  \fbt{and all spin-spin correlations in $\rho^\mathrm{S}_{AB}$ fall below the threshold where their quantum coherences are indistinguishable from a classical mixture.  In the 2s limit $N^\mathrm{S}=0$ exactly corresponds to no remaining spin entanglement \cite{Horodecki_1996}\khcax{. Whereas}\khca{;} in the dimer-dimer setting there can be remaining bound entangled states that are not detected by $N^\mathrm{S}$. From the non\khca{-}existence of $N^\mathrm{S}$ at large distances, we conclude that the fact that $N^\mathrm{F}$ is nonzero for $d_x>1$ is due to the presence of other \khcax{non-diagonal}\khca{off-diagonal} elements of the reduced density matrix, which take into account other, non spin-spin correlations. These non spin-spin correlators} \khcax{are decaying}\khca{decay} exponentially in both the \emph{BI} and \emph{MI} phases. Thus we can suggest the following general asymptotic behavior of $N^\mathrm{F}$ at large distances, expressed in terms of spin-spin correlations ${\cal C}_\mathrm{S}$ and correlations in \khca{the} charge (gapped) sector ${\cal C}_\mathrm{Q}$:
\begin{eqnarray}
    N^\mathrm{F} (d)= F(C_0 + {\cal C}_\mathrm{S} (d), {\cal C}_\mathrm{Q} (d) ) {\cal C}_\mathrm{Q} (d),\label{eq:NF_MI_BI}
\end{eqnarray}
where $F$ is some function of \khcax{both correlations in charge and spin sector}\khca{the correlations in the charge and spin sectors}. 
Since $N^\mathrm{S}=0$ at large distances while $N^\mathrm{F}$ is not, we know that $F$ can not be directly proportional to spin-spin correlations and the charge sector alone must be enough to support a non-zero $N^\mathrm{F}$. Furthermore, any contribution to $F$ must include diagonal elements of the reduced density matrix \cite{PhysRevB.110.075115}, these include disconnected correlators converging to a constants $C_0$ in the limit of large separation $d$. 

In principle, the prefactor in front of ${\cal C}_\mathrm{Q}$ is still dependent on spin correlations, which change non-analytically at large distances. However, due to the aforementioned structure of the $F$ prefactor, even a divergent correlation length for the spin correlations with a switch from polynomial to exponential behavior \khcax{can not}\khca{cannot} lead to the appearance of terms in the logarithmic derivative of $N^\mathrm{F}$ which are growing with the distance $d$ (the peaks seen \khcax{on Figures ~}\khca{in Figs.~}\ref{fig:dimer_dimer_NF_MI_BI} and \ref{fig:2s_NF_NI_BI}). Also, the best fit results from Table \ref{tab:fits_MI_BI} do not show large (up to several times) differences in correlation length values between fits before and after the transition, with the opposite picture being typical for the \emph{SM-BI} transition, shown in Table \ref{tab:fits_SM_BI}. Thus we conclude that the peaks in the logarithmic derivative \khcax{can not}\khca{cannot} be directly explained by the divergent spin correlation length. Instead, it is the indirect influence of critical spin correlations onto the charge sector which is responsible for potential divergences in the logarithmic derivative of $N^\mathrm{F}$. This statement can be formalized once we write down the logarithmic derivative of $N^\mathrm{F}$ according to \khca{Eq.~\eqref{eq:NF_MI_BI}}
\begin{align}
    \frac{\partial \log N^\mathrm{F} (d)}{\partial t'} = & \frac{1}{F} \left( \partial_1 F \left( \frac{\partial C_0}{\partial t'} + \frac{\partial {\cal C}_\mathrm{S}}{\partial t'} \right)  +  \partial_2 F  \frac{\partial {\cal C}_\mathrm{Q}}{\partial t'}  \right)  \nonumber \\
    & + \frac{\partial \log {\cal C}_\mathrm{Q}}{\partial t'},\label{eq:NF_MI_BI_der}
\end{align}
 
where $\partial_1 F$ and $\partial_2 F$ are partial derivatives of $F$ with respect to the first and the second parameter. Now, assuming  ${\cal C}_\mathrm{Q} \sim e^{-d/\xi(t')}$ and  ${\cal C}_\mathrm{S} \sim e^{-d/\xi_\mathrm{S}(t')}$, the second term leads to 
\begin{eqnarray}
    \frac{\partial \log {\cal C}_\mathrm{Q}}{\partial t'} = \frac{d}{\xi^2} \frac{\partial \xi}{\partial t'},\label{eq:NF_MI_BI_dominant_der}
\end{eqnarray}
while the first term only contains the derivatives of correlation lengths multiplied by exponentials:
\begin{eqnarray}
    \frac{\partial {\cal C}_\mathrm{Q}}{\partial t'} = \frac{d}{\xi^2}\, e^{-d/\xi} \frac{\partial \xi}{\partial t'}, \\
    \frac{\partial {\cal C}_\mathrm{S}}{\partial t'} = \frac{d}{\xi_\mathrm{S}^2}\, e^{-d/\xi_\mathrm{S}} \frac{\partial \xi_\mathrm{S}}{\partial t'}.
\end{eqnarray}
 These derivatives are also combined with \khcax{expression}\khca{expressions} which are constant in the large $d$ limit, like $\frac{\partial C_0}{\partial t'}$. Thus only the second term, originating from charge sector correlations, \khcax{can provide linear asymptotic in $d$ potentially}\khca{can potentially provide a term linear in $d$}, leading to the distinct peak in the first or second logarithmic derivatives. However, since the charge degrees of freedom are gapped and hence not critical, we need an additional explanation why the charge correlation length $\xi$ still picks up the transition.  
 
 Upon  analysis of the  RDM, the correlation  functions that are 
 responsible for the  observed  fermion negativity  read $\left< \hat{c}^{\dagger}_{x,\sigma} \hat{n}_x \hat{n}_y \hat{c}^{}_{y,\sigma}  \right>$. In what follows, we  show  that the charge fluctuations  pick  up the criticality  and will then assume a mean field like  decoupling between the  single particle  Green function and the charge fluctuations. 
 Let $S_{\phi}  =  \int d^3 x  (\partial_\mu \bm{\phi})^2   - m_{\phi} \bm{\phi}^2  + \lambda \bm{\phi}^4  $.  Here, $m_{\phi}$  drives  the  transition  and  $\bm{\phi}$ corresponds to the three\khca{-}component  magnetic order parameter. Let $\rho(x) $ correspond to the massive  charge field. The symmetry-allowed coupling between the  charge and  spin degrees of freedom reads:  $S_{\rho, \phi} =  g \int d^3 x \bm{\phi}^2 \rho $. Clearly\khca{,}  since  $\rho$ is  massive\khca{,} this coupling is irrelevant at the  O(3) quantum critical point. However,  the  charge-charge correlations  will pick  up  the  scaling  dimension of the  energy operator.  In particular, we  expect: 
\begin{equation}
        \langle \rho(x)\rho(y) \rangle   \simeq G_{\rho}(x-y) \langle \bm{\phi}^2(x) \bm{\phi}^2(y)   \rangle. 
\end{equation} 
Here $G_{\rho}(x-y)$ is the propagator of  the massive  charge  fluctuations,   and  in the vicinity of  criticality we expect: 
\begin{equation}
     \langle \bm{\phi}^2(x) \bm{\phi}^2(y)   \rangle \propto  \frac{1}{|x-y|^{2 \Delta_\epsilon}}  e^{-|x-y|/\xi(t')}
\end{equation}
Here $\Delta_\epsilon$ is the scaling dimension of  the  energy operator  and $\xi \propto |t'-t'_\mathrm{c}|^{-\nu}$  the  correlation length  that diverges at criticality. For the O(3) transition, the  correlation length  exponent 
reads $\nu \simeq 0.711$ \cite{Vicari02} such that the log derivative of the fermionic negativity should  show  a divergence as a function of  distance, since   $\frac{1}{\xi^2} \frac{\partial \xi}{\partial t'} \sim |t'-t'_\mathrm{c}|^{\nu-1}$ diverges at the transition point.  This behavior is  supported by the data of Fig.~\ref{fig:dimer_dimer_NF_MI_BI} \fbc{and Fig.~\ref{fig:2s_NF_NI_BI}}. 

\subsection{SM-MI transition}
\begin{figure}
    \centering
    \includegraphics[width=\linewidth]{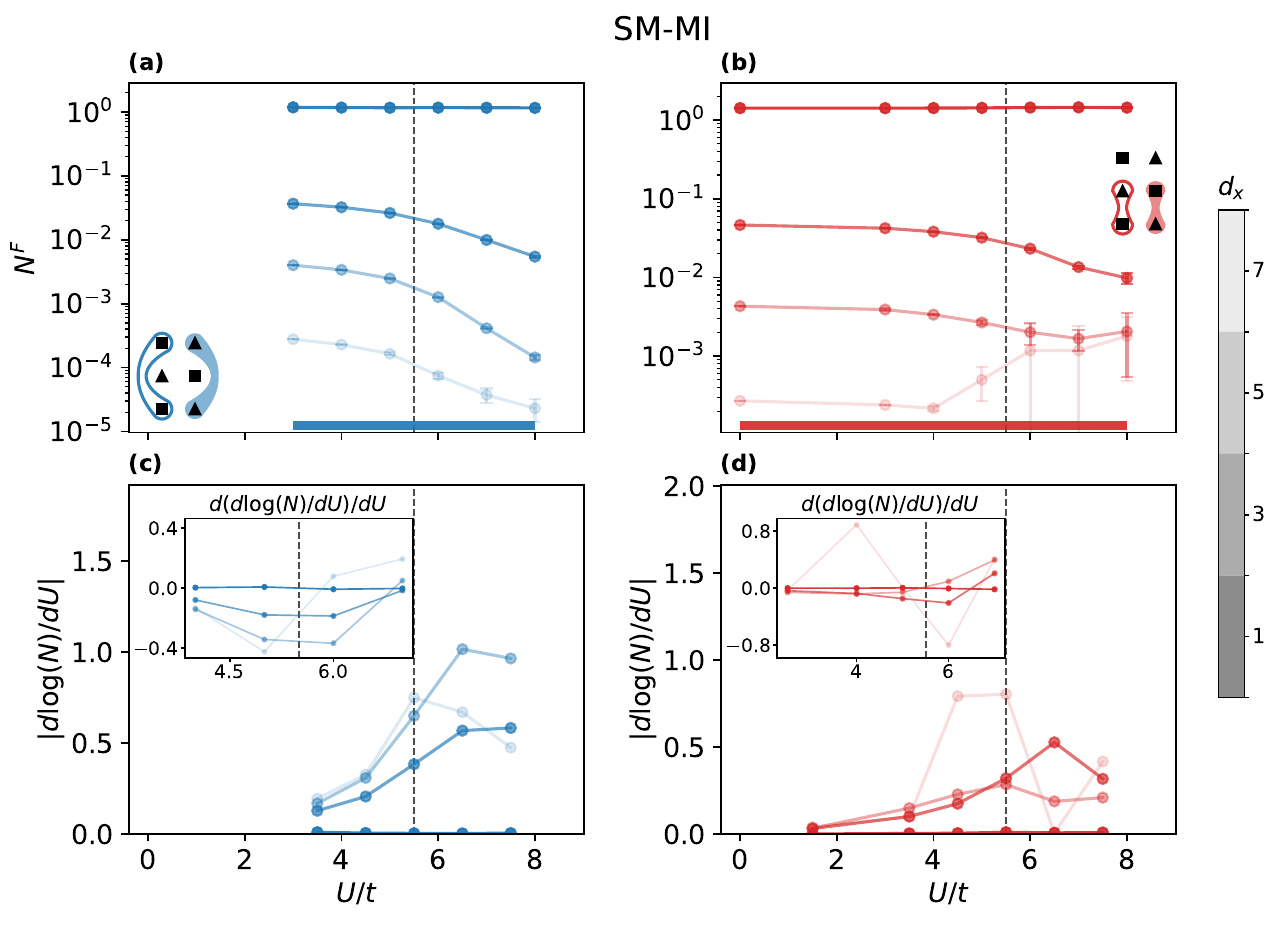}
    \caption{Same as Figs.~\ref{fig:2s_NF_SM_BI}-\ref{fig:dimer_dimer_NF_MI_BI} but for dimer-dimer entanglement across the \emph{SM}-\emph{MI} transition.
    }
    \label{fig:dimer_dimer_NF_SM_MI}
\end{figure}

\begin{figure}
    \centering
    \includegraphics[width=\linewidth]{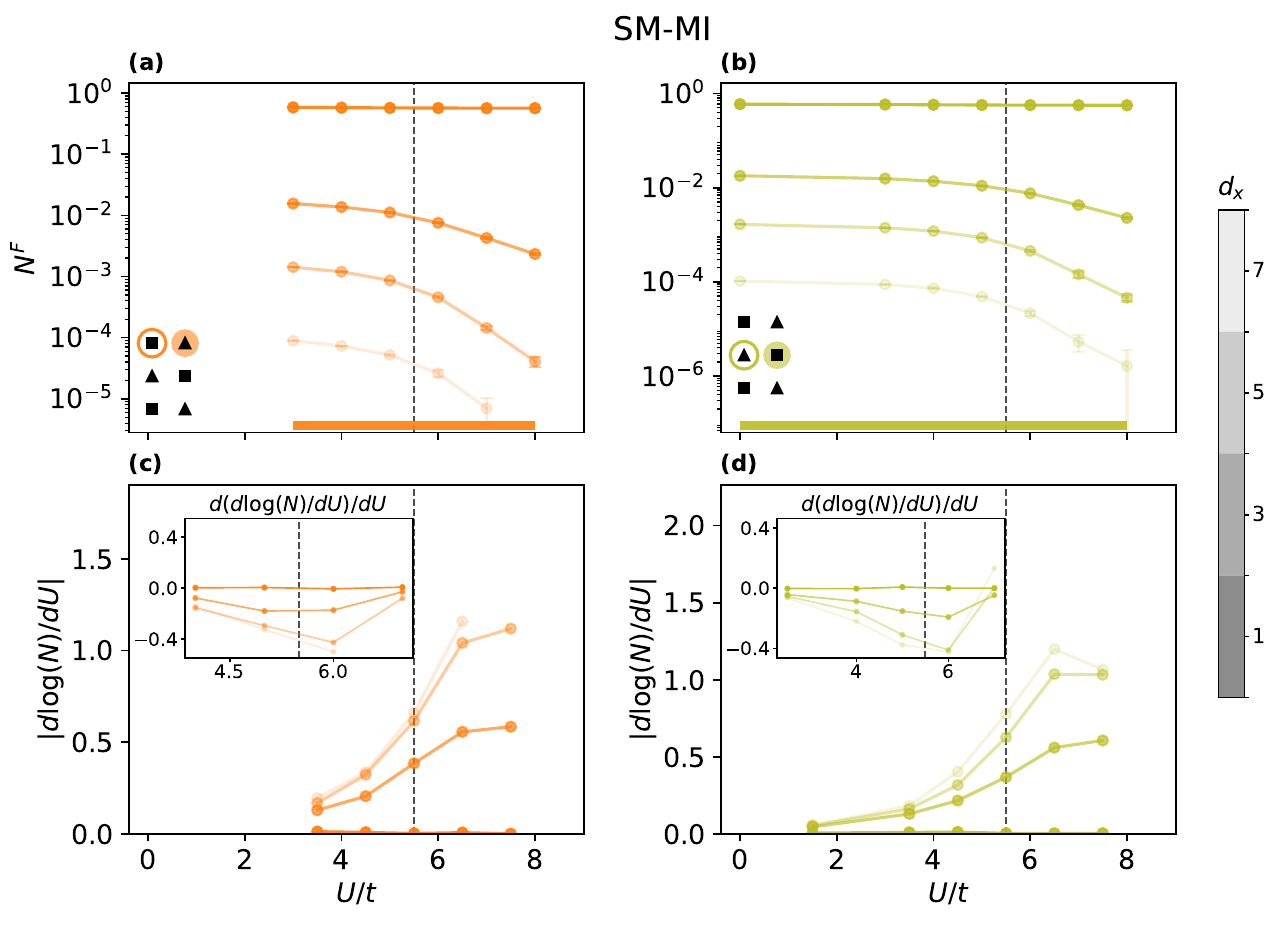}
    \caption{Same as Fig.~\ref{fig:2s_NF_NI_BI} but for the \emph{SM}-\emph{MI} transition.}
    \label{fig:2s_NF_SM_MI}
\end{figure}

\begin{figure}
    \centering
    \includegraphics[width=\linewidth]{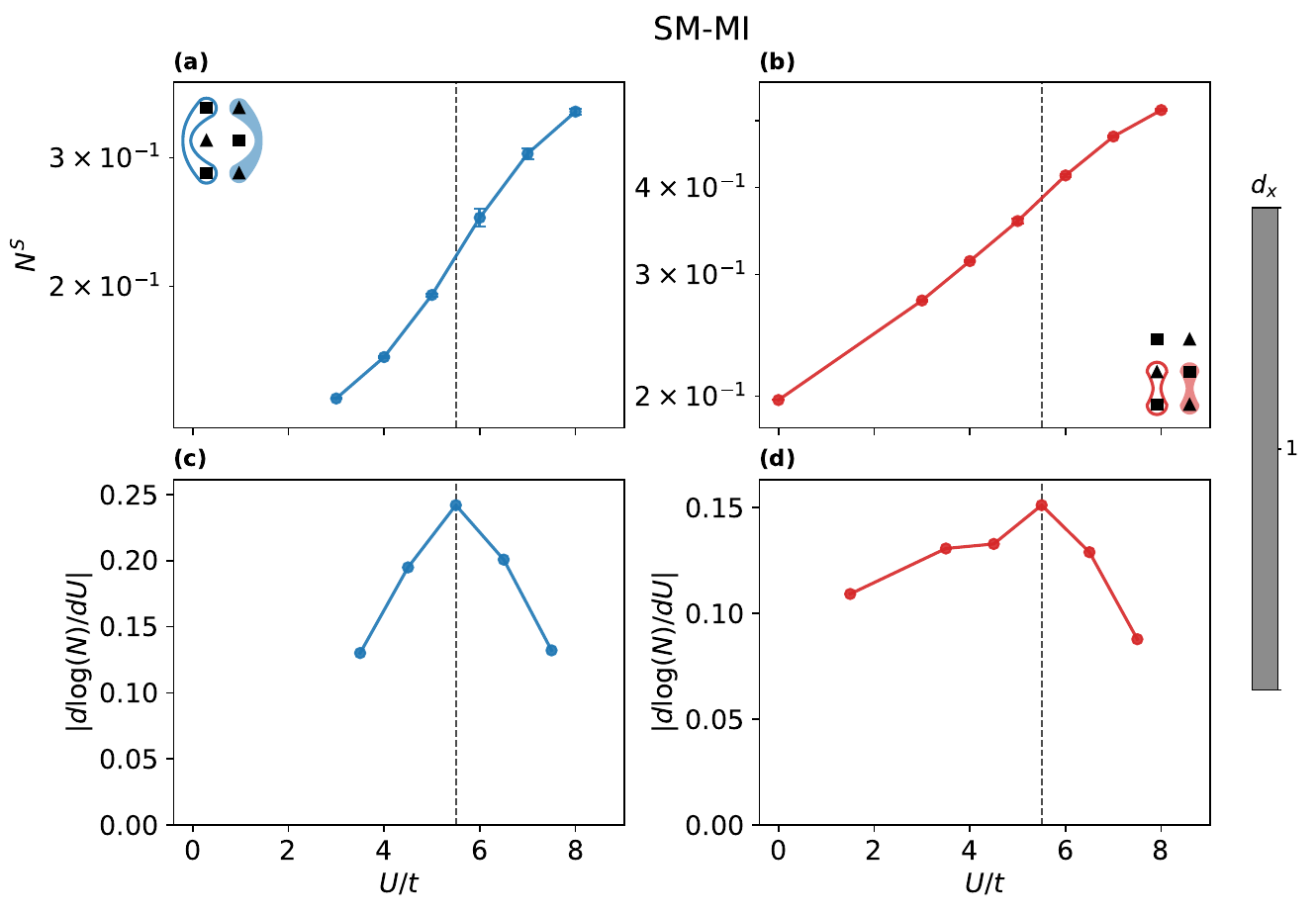}
    \caption{Same as Fig.~\ref{fig:dimer_dimer_NF_SM_MI} but now with spin entanglement $N^\mathrm{S}$ across the \emph{SM}-\emph{MI} transition for the dimer-dimer geometry.}
    \label{fig:dimer_dimer_NS_SM_MI}
\end{figure}

\begin{figure}
    \centering
    \includegraphics[width=\linewidth]{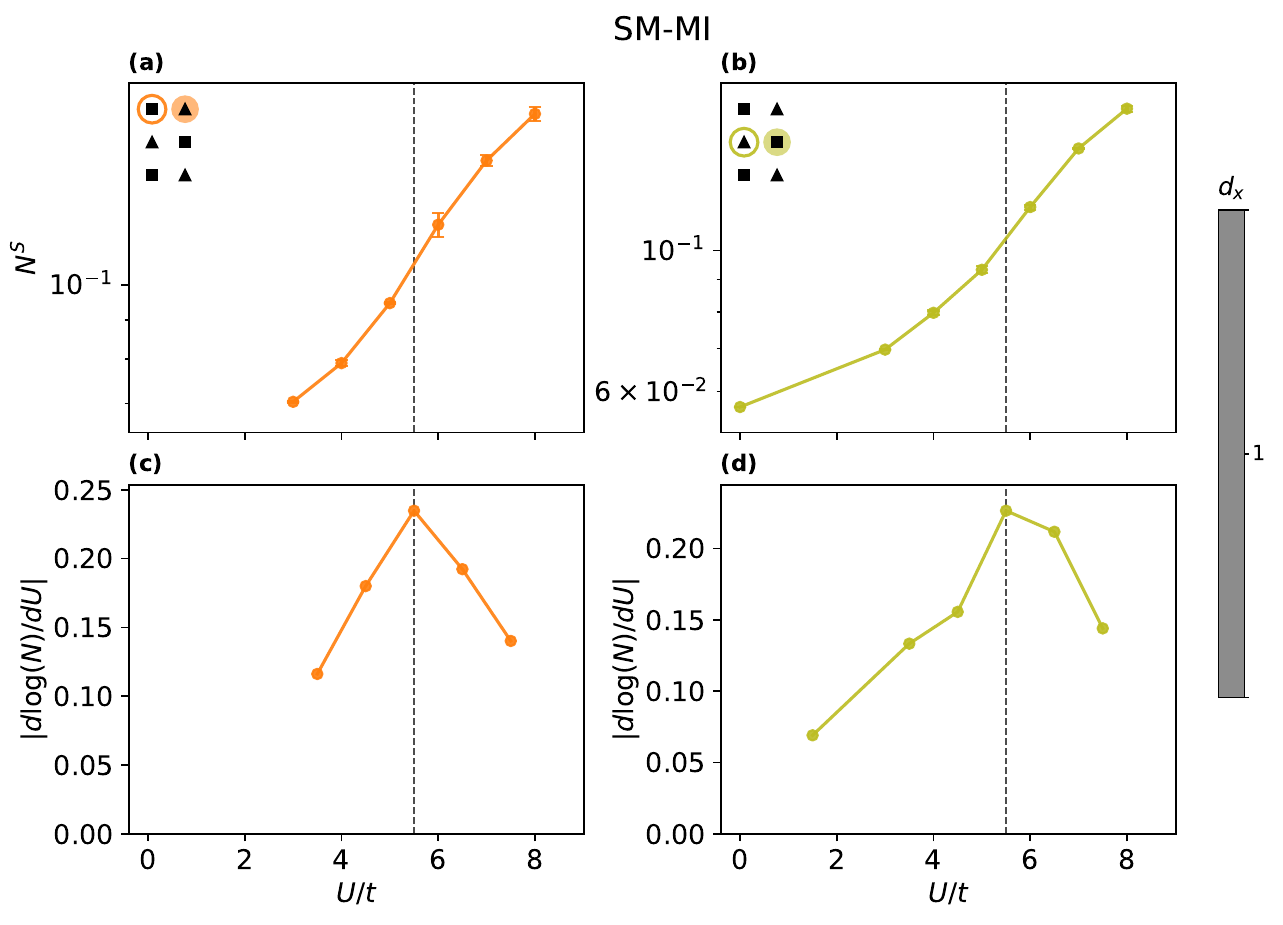}
    \caption{2s spin entanglement $N^\mathrm{S}$ and logarithmic derivatives for the \emph{SM}-\emph{MI} transition.}
    \label{fig:2s_NS_SM_MI}
\end{figure}

\begin{table}[t]
\centering
\caption{Fits of $N^\mathrm{F}(d) = \kappa\, d^{-\epsilon} \fbc{e^{-d/\xi}}$ for the \emph{SM}-\emph{MI} transition.
$^\dagger$Unphysical fit ($\epsilon < 0$ or $\fbc{\xi} < 0$).}
\label{tab:fits_SM_MI}
\begin{tabular}{l c | c c}
partition & $U/t$ & $\epsilon$ & $\fbc{\xi}$  \\
\hline
\hline
\textcolor{fitblue}{blue}     & 3.0 & 1.23 & 1.03 \\
                              & 6.0 & 1.91 & 1.03 \\
\hline
\textcolor{fitorange}{orange} & 3.0 & 1.27 & 0.97 \\
                              & 6.0 & 2.03 & 1.01 \\
\hline
\textcolor{fitgreen}{green}   & 3.0 & $-2.51^\dagger$ & $0.63^\dagger$ \\
                              & 6.0 & $-1.39^\dagger$ & $0.66^\dagger$ \\
\hline
\textcolor{fitred}{red}       & 3.0 & 1.10 & 0.93 \\
                              & 6.0 & $4.42^\dagger$ & $-4.69^\dagger$ \\
\hline
\textcolor{fitkhaki}{khaki}   & 3.0 & 1.26 & 0.96 \\
                              & 6.0 & 1.80 & 0.91 \\
\end{tabular}
\end{table}

As mentioned in the introduction, this transition belongs to the Gross-Neveu-Heisenberg universality class, in which  the fermion mass is generated through  spontaneous symmetry breaking, in this case  antiferromagnetism.  In the large-N limit this transition can be accounted for by a  mean-field  theory, in which the  single\khca{-}particle Green's  function scales 
as 
\begin{align}
{\cal C}(d) \sim d ^{-\alpha} e^{-d/\xi(U)}
\label{eq:G_polynom_exp_two}
\end{align}
in the  vicinity of the  quantum phase transition.  With $\xi(U) \sim  | U -U_\mathrm{c}|^{-\nu} $  with $\nu$ 
the correlation length exponent,  the log derivative with respect to $U$ of ${\cal C}(d)$ reads:
\begin{align}
\frac{\partial \log {\cal C}(d)}{\partial U} 
\sim d |U - U_\mathrm{c}|^{\nu-1}
\end{align}
Using the mean-field relation between the fermionic negativity and the fermion correlation function of  
Eq.~\eqref{eq_NF_from_G},  the numerical value of the correlation length exponent  will dictate the scaling 
behavior of this quantity.  Recent large-scale auxiliary-field quantum Monte Carlo simulations  suggest 
$\nu = 1.11(9)$ \cite{Wang26}. This  implies that the first logarithmic derivative will not show a divergent behavior but that the second  will. Upon inspection,  one can indeed observe  that the numerical data in Fig.~\ref{fig:dimer_dimer_NF_SM_MI} (dimer-dimer) and Fig.~\ref{fig:2s_NF_SM_MI} (2s) exhibit this behavior.  In particular  we observe a clearer peak in the second logarithmic derivative  than in the first one.

Since the transition is, however, characterized by the localization of electrons, 
it is most natural to look at the spin entanglement measured by $N^\mathrm{S}$.
In Fig.~\ref{fig:dimer_dimer_NS_SM_MI} $N^\mathrm{S}$ between dimers unveils again a clear
maximum in its logarithmic derivative.
Due to the sudden death of spin entanglement, no comparison to larger distances is possible. This also holds for the 2s data in Fig.~\ref{fig:2s_NS_SM_MI}.

\section{Summary}
We have used auxiliary-field quantum Monte Carlo methods to compute reduced density matrices for a \textit{small} number of sites from correlation functions. Given the reduced density matrix of two disconnected subregions separated by a distance $d$, we ask whether entanglement measures can detect various quantum phase transitions. These measures include the fermionic negativity and the spin entanglement.
To this end, 
Osterloh {\it et al.} have shown 
that the system-size scaling of  two-spin entanglement can be used to detect quantum phase transitions 
\cite{osterloh_scaling_2002}. Our work extends this
idea to interacting fermionic systems for which large system sizes are inaccessible. In these systems, we instead consider two small partitions subsystems and increase the distance between them. Specifically, we study
the $t'$-$\pi$-flux Hubbard model as a representative model exhibiting the \emph{SM}-\emph{BI}, \emph{SM}-\emph{MI}, and \emph{MI}-\emph{BI} transitions, which belong to the Gaussian, 2+1D Gross-Neveu-Heisenberg \cite{Herbut09a}, and 2+1D O(3) Wilson-Fisher \cite{Vicari02} universality classes, respectively.

In the aforementioned model, we observe the sudden death of spin entanglement at distances greater than the lattice spacing. The quantum Fisher information \cite{hauke_measuring_2016}, obtained from the spin susceptibility, provides a lower bound on multipartite entanglement. For quantum phase transitions with an anomalous-dimension exponent below unity, this implies that multipartite entanglement diverges at the critical point. This is the case for the O(3) Wilson-Fisher universality class, showing that entanglement manifests itself in various forms.

While spin entanglement exhibits sudden death at distances greater than the lattice spacing, the fermionic negativity remains nonzero. The key to identifying distinct signatures of the transitions is to consider the logarithmic derivative of the fermionic negativity with respect to the tuning parameter. For all the transitions considered, this quantity scales as $|g-g_\mathrm{c}|^{\nu - 1}$, where $g$ is the tuning parameter, $g_\mathrm{c}$ is the critical point, and $\nu$ is the correlation-length exponent.

For the Gaussian (\emph{SM}-\emph{BI}) transition, $\nu=1$, so the first logarithmic derivative develops a jump and the second a peak, as observed above. For the O(3) Wilson-Fisher (\emph{MI}-\emph{BI}) transition, $\nu<1$ implies a divergence at the critical point. For the 2+1D Gross-Neveu-Heisenberg transition (\emph{SM}-\emph{MI}), recent computations \cite{Wang26} suggest that $2 > \nu > 1$, so the singularity is expected to be absent in the first logarithmic derivative but present in the second.

The fermionic negativity considered here is the same  diagnostic tool  across the variety of transitions exhibited by the $t'$-$\pi$-flux Hubbard model. Since it captures the correlation-length exponent, or equivalently the scaling dimension of the relevant operator that defines the critical point, we conjecture that it provides a universal tool for detecting quantum phase transitions, even those that do not involve a conventional order parameter. Examples include Kondo-destruction transitions \cite{Si01,mazza_quantum_2024,Pan25}.

\begin{acknowledgements}
We thank A.~Kauch for discussions. 
{We acknowledge assistance from Claude (Anthropic) 
with plotting, mathematics,  writing, and some formal checks.}
This work has been supported in part by the Austrian Science Fund(FWF) through the FWF Spezialforschungsbereich (SFB) QM\&S project DOI 10.55776/F86.
The authors gratefully acknowledge the Gauss Centre for Supercomputing e.V. (www.gauss-centre.eu) for funding this project by providing computing time on the GCS Supercomputer JUWELS  ~\cite{JUWELS} at the Jülich Supercomputing Centre (JSC) and on the GCS Supercomputer SUPERMUC-NG at the Leibniz Supercomputing Centre (www.lrz.de, project number pn73xu).
The Flatiron Institute is a division of the Simons Foundation.
\end{acknowledgements}

\section*{DATA AVAILABILITY}
The data that supports the findings of this article will be made publicly available upon publication.


\clearpage

\appendix
\section{Non\kh{-}entanglement based detection of the phase transition}\label{Appendix_phase_transitions}
\begin{figure}
    \includegraphics[width=1\linewidth]{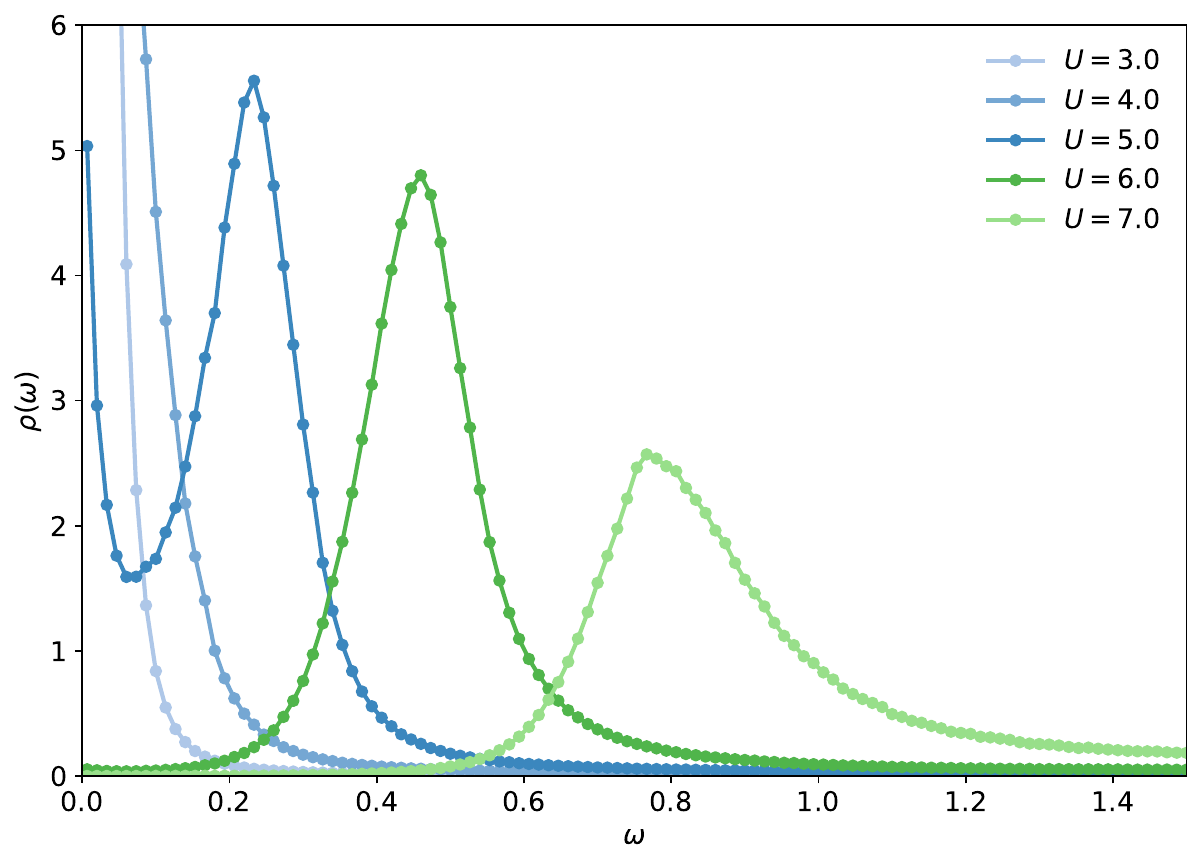}
    \caption{Single\khca{-}particle spectral functions at \khca{the} Dirac point.}
    \label{fig:spectra_SM_MI}
\end{figure}

In this Appendix we briefly reproduce previous results on the phase diagram of \khca{the} $t'$-$\pi$-flux model to demonstrate consistency of our QMC simulations with previous works. First, we check the pure $\pi$-flux model, which corresponds to $t'=t$, and compute the single-particle spectral function at the Dirac point. Results are shown in Fig.~\ref{fig:spectra_SM_MI}. One can see that the gap appears at interaction strength between $U=5$ and $U=6$, which is consistent with the previous estimate $U_\mathrm{c}\approx5.5$.

\begin{figure}
    \includegraphics[width=1\linewidth]{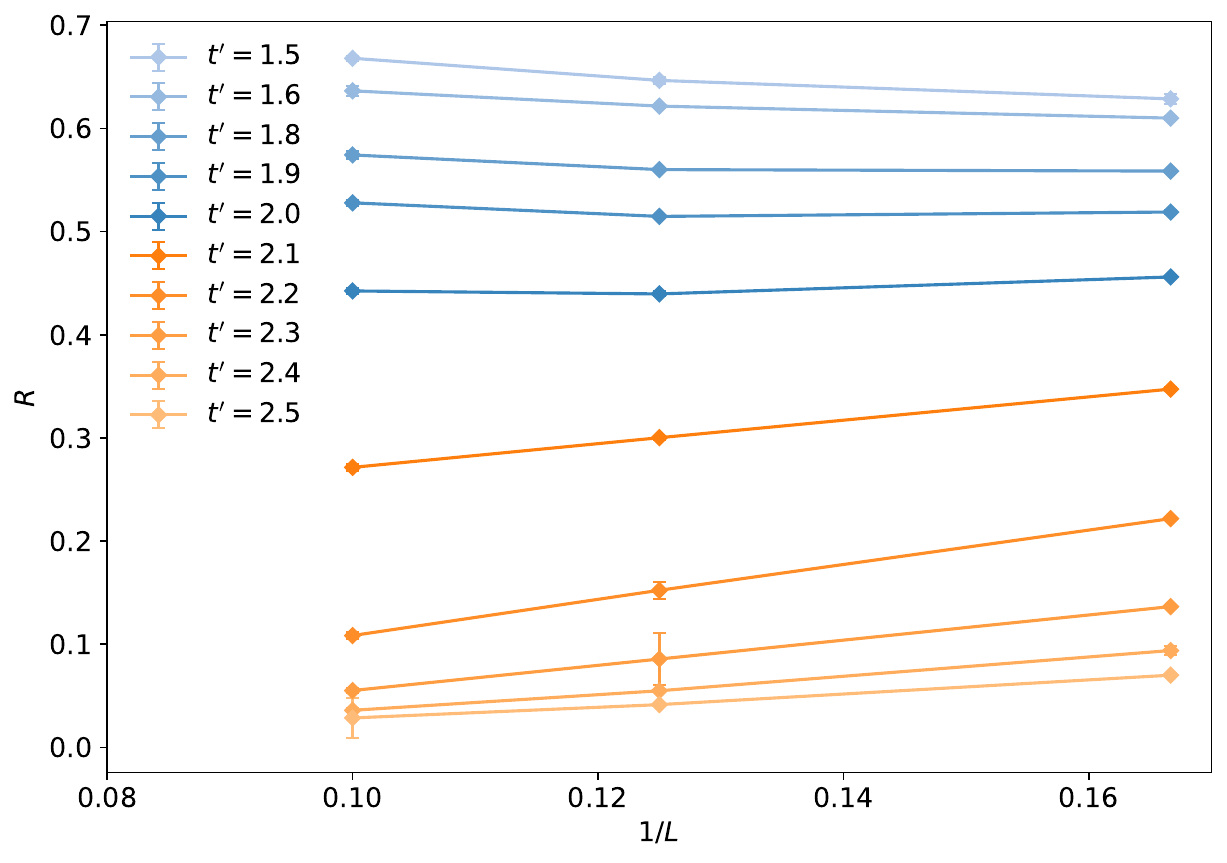}
    \caption{\khcax{R-ratio}\khca{Correlation ratio $R$:} scaling with system size to identify the phase transition between Mott AFM insulator and band insulator phase.}
    \label{fig:R_ratio}
\end{figure}

Next we check the location of the \emph{MI}-\emph{BI} transition (Fig.  \ref{fig:R_ratio}). We use the standard method of detecting the appearance of ordering using the correlation ratio: 
\begin{align}
R=1- \frac{S(\vec Q+\Delta \vec q)}{S(\vec Q)},
\end{align}
where $\vec Q$ is the ordering vector in the BZ for spin correlations, $S(\vec q)$ is the spin structure factor and $\Delta \vec q$ is the minimal shift for lattice momentum at a given lattice size. 
The correlation ratio increases with increased lattice size for $t'<2.0$ and decreases for $t'>2.1$ which corresponds to the transition between \emph{MI} and \emph{BI} phase at $t'$ value between these two points, in full agreement with previous QMC results reported in \khca{Ref.~}\cite{Ixert_piflux_2014}. 

\section{Connection between superselection rules and spin negativities}\label{SSR_vs_spin_resolved}

In this Appendix we show that
\begin{equation}
    0 \leq N^{\textnormal{S}} \leq N^{\textnormal{NSSR}} \leq N^{\textnormal{PSSR}}. 
\label{eq:s<NSSR<PSSR}
\end{equation}

This implies that whenever the spin negativity detects entanglement 
there is entanglement in the NSSR sense too.
\kh{Here, t}he three projections are refinements of each 
other\kh{: the \kh{parity} PSSR density matrix erases the matrix elements
connecting different parity subspaces.} When we construct the \kh{charge} 
NSSR superselected density matrix, we divide the subspaces into smaller ones
according to the particle numbers of the subsystem\kh{s $A$ and $B$;} and 
if we introduce the site\kh{-}wise particle\kh{-}number\kh{-}projected 
density matrix, we divide the subspaces even further \kh{
with fixed particle numbers on all sites} and delete the matrix elements 
connecting the subspaces.
So the density matrices can be constructed by repeated projections.
\begin{align}
   \rho^{\textnormal{PSSR}} &= \sum_{\Pi_A,\Pi_B} P_{\Pi_A} \otimes P_{\Pi_B} \enspace  \rho \enspace P_{\Pi_A} \otimes P_{\Pi_B} \\
   \rho^{\textnormal{NSSR}} &= \sum_{ N_{A}, N_{B}  } P_{N_A} \otimes  P_{N_B} \rho^\text{PSSR}  P_{N_A} \otimes  P_{N_B}\\
\rho_{AB}^{\textrm{S}} &= \nonumber \\
&\sum_{{N_{i},N_{j}}}  \prod_{i\in A} P_{N_i} \otimes \prod_{j \in B} P_{N_j} \;  \rho^\text{NSSR}_{AB} \;  \prod_{i\in A} P_{N_i}  \otimes\prod_{j\in B} P_{N_j},
\end{align}
where $\Pi_A$ and $\Pi_B$ label the parity sectors of the two subsystems,
$N_A$ and $N_B$ similarly label the subsystem charge sectors and $N_i$ and
$N_j$ are single-site charges.
Applying this to each refinement step yields 
All three projections commute with the partial transpose. The projections \kh{restrict} the 
parity of the particle number (PSSR) or the particle number \kh{itself}  (NSSR) of \kh{the} subsystems\kh{;} or the particle number of \kh{all} sites \kh{(S)}.
Now the partial transpose just interchanges the  state indices of one subsystem \kh{(say $B$)} between the left and the right, so if the conditions were met before the partial transpose, they will be met after as well.

\begin{equation}
    \sum_r P_r \rho^{T_B} P_r =  \left( \sum_r P_r \rho P_r \right)^{T_B}
\end{equation}
where $P_r$ stands for the projectors defined above.

To arrive the inequalities of Eq.~\eqref{eq:s<NSSR<PSSR}
we represent the refinement projection an average of unitary conjugations, \begin{equation}
    \sum_{r=1}^nP_rXP_r=\frac1n\sum_{k=1}^nU_kXU_k^\dagger
\end{equation} with $U_k=\sum_re^{2\pi ikr/n}P_r$. 
Here we label the projectors $1 \dots n$, where $n$ is the number of invariant subspaces. In the case of parity $n=2^2=4$, in the case of NSSR it is $4 |A| |B|$, where $|A|$ ( $|B|$)  are the number of sites in the $A$ ($B$)  subsystem, and in the case of the  projection to the local particle numbers it is $4^{|A| + |B|}$
By the triangle inequality, and because unitary conjugation preserves the trace norm, this projection cannot increase the trace norm. Since the projections commute with the partial transpose, we have
\begin{equation}
  \|(\sum_rP_r\rho P_r)^{T_B}\|_\mathrm{tr}=\|\sum_rP_r\rho^{T_B}P_r\|_\mathrm{tr}\le\|\rho^{T_B}\|_\mathrm{tr},  
\end{equation}
applying this to each refinement step yields Eq.~\eqref{eq:s<NSSR<PSSR}.

\section{Mutual Information}
\label{Appendix_I}
\begin{figure}
    \centering
    \includegraphics[width=\linewidth]{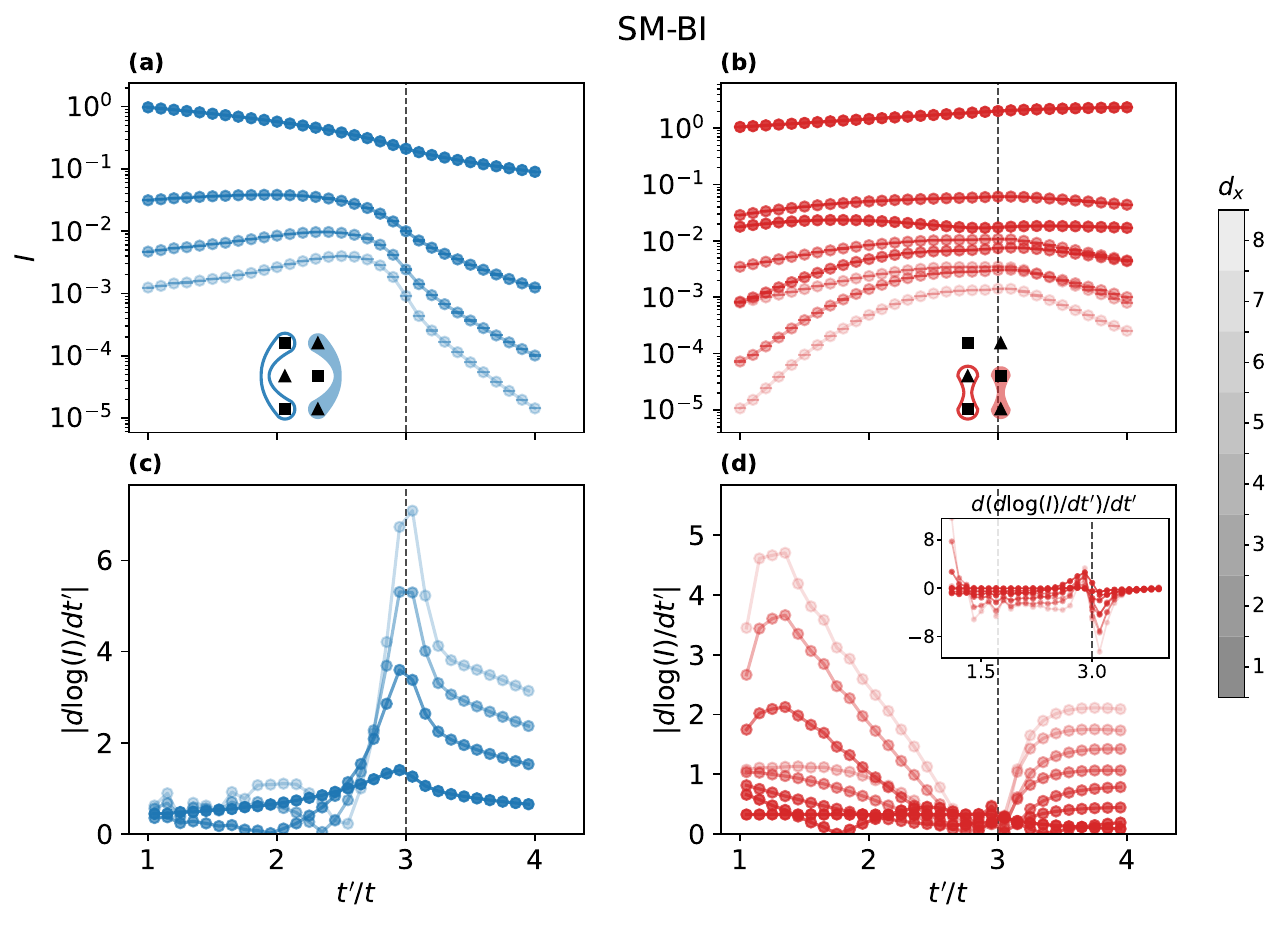}
    \caption{Dimer-dimer mutual information $I$ for the \emph{SM}-\emph{BI} transition. 
    }
    \label{fig:dimer_dimer_I_SM_BI}
\end{figure}

\begin{figure}
    \centering
    \includegraphics[width=\linewidth]{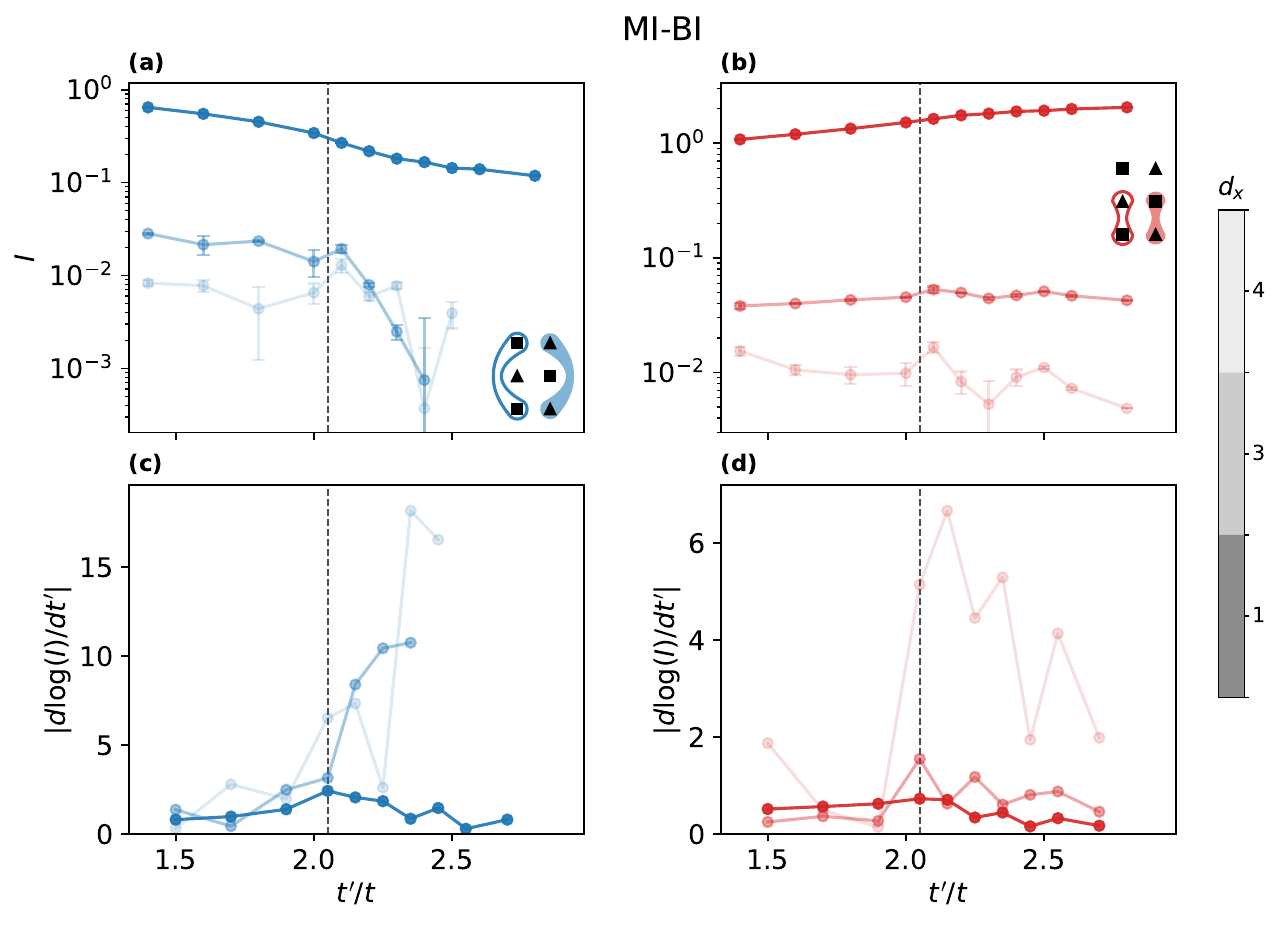}
    \caption{Same as Fig.~\ref{fig:dimer_dimer_I_SM_BI} but for the \emph{MI}-\emph{BI} transition.}
    \label{fig:dimer_dimer_I_MI_BI}
\end{figure}

\begin{figure}
    \centering
    \includegraphics[width=\linewidth]{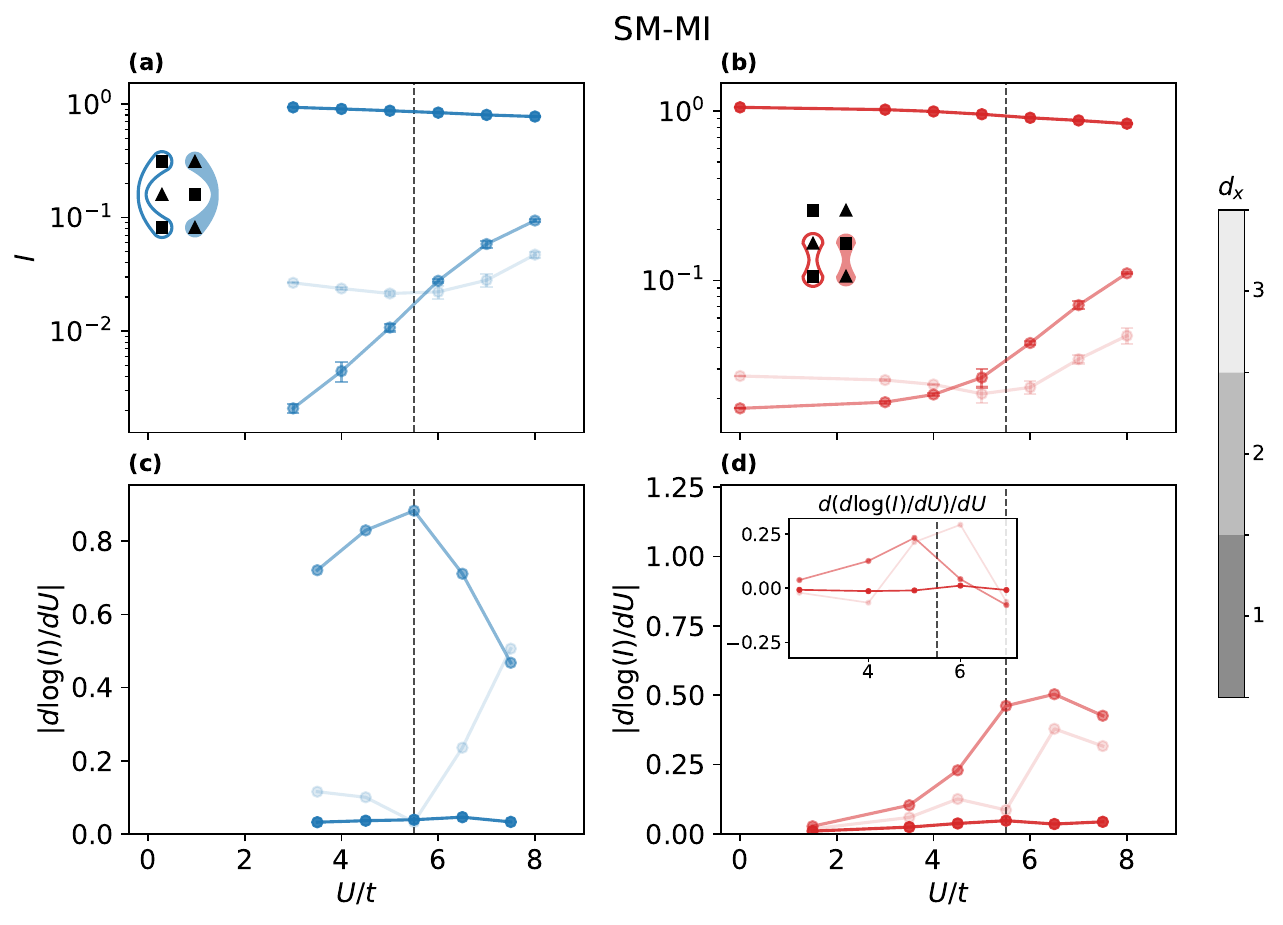}
    \caption{Same as Fig.~\ref{fig:dimer_dimer_I_SM_BI} but for the \emph{SM}-\emph{MI} transition.}
    \label{fig:dimer_dimer_I_SM_MI}
\end{figure}

\begin{figure}
    \centering
    \includegraphics[width=\linewidth]{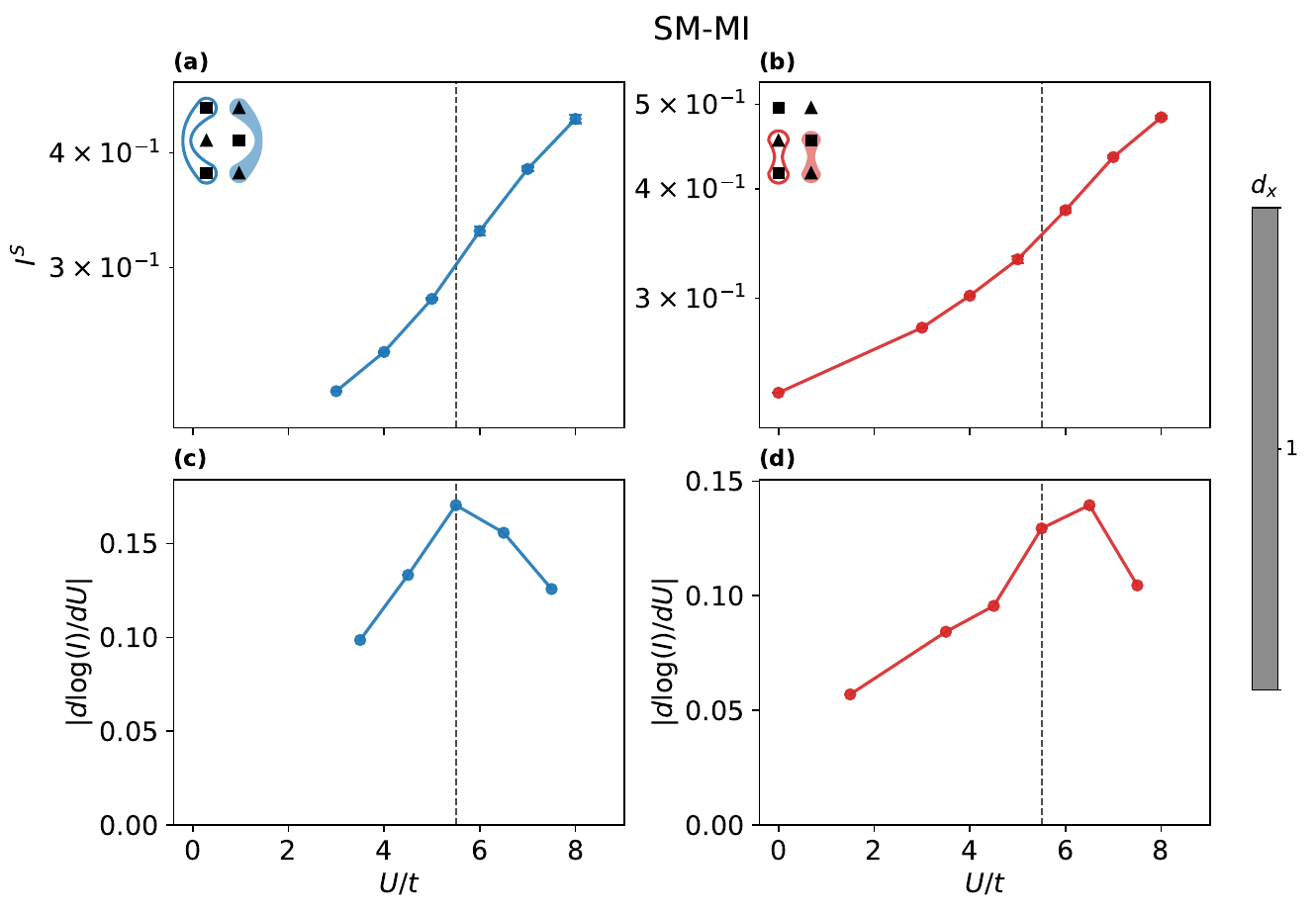}
    \caption{Dimer-dimer spin mutual information $I^\mathrm{S}$ for the \emph{SM}-\emph{MI} transition.}
    \label{fig:dimer_dimer_IS_SM_MI}
\end{figure}

We further complement the results for the negativity by comparing to the mutual 
information, a measure of all correlations in a system \cite{zurek_information_1983,barnett_entropy_1989,amico_entanglement_2008,Groisman_2005,Wolf_2008}. 
The mutual information is defined as
\begin{align}
    I_{AB} & \equiv \textrm{tr}[\rho_{AB}\log\rho_{AB}] - \textrm{tr}[\rho_{A}\log\rho_{A}] - \textrm{tr}[\rho_{B}\log\rho_{B}] \nonumber\\
    & = \textrm{tr}[\rho_{AB}\log\rho_{AB}] - 2 \textrm{tr}[\rho_{A}\log\rho_{A}].
\end{align}
The last equality holds 
\kh{if $A$ and $B$ are equivalent by symmetry}. Fermionic signs need to be
taken into account for computing the subsystem density matrices. 
The mutual information measures the difference between the
von Neumann entropy of the system under the assumption of \kh{subsystems $A$ and $B$} being completely 
uncorrelated and the actual realized system\kh{. H}ence, 
if $I=0$ the system is not correlated and consequently has no entanglement.
In \kh{contrast} to the negativity, the mutual information captures classical 
correlations, quantum correlations beyond entanglement, and entanglement 
\cite{henderson_classical_2001}. Note that for a pure state $\rho_{AB}$ of two subsystems is mixed also at $T=0$

Fig.~\ref{fig:dimer_dimer_I_SM_BI} through Fig.~\ref{fig:dimer_dimer_I_SM_MI}
show the mutual information for the dimer-dimer pairings across the three different 
transitions and the derivatives thereof. The results are in good agreement with the
previous findings from the fermionic negativity. Note, however, that unlike the negativity, we do not correct for the slightly noisy trace here which can lead to significant fluctuations. While the transitions are not 
directly visible in the mutual information, the appropriate logarithmic 
derivatives show a sharp feature at large distances. Again, only considering the 
mutual information obtained from the spin-filtered density matrix $\rho^{\mathrm S}$ 
shown in Fig.~\ref{fig:dimer_dimer_IS_SM_MI} can be helpful to make the 
interaction\khca{-}driven \emph{SM}-\emph{MI} transition apparent at short distances.
From a qualitative comparison with the results for the negativity,
it also becomes clear that the features of the phase transitions which can be observed in the mutual information are (at least partially) due to entanglement and not 
explainable solely through classical correlations since they already exist in the 
negativity results.

\bibliography{main_references.bib}

\end{document}